\documentclass[twocolumn,prb]{revtex4-1}

\usepackage{graphicx}
\usepackage{epsfig}
\usepackage{amsmath, amssymb}
\usepackage{verbatim}
\usepackage{bm}

\def\sgn{{\rm sgn}}

\begin{document}

\title{Dissipationless Phonon Hall Viscosity}
\author{Maissam Barkeshli}
\author{Suk Bum Chung}
\author{Xiao-Liang Qi}
\affiliation{Department of Physics, Stanford University, Stanford, CA 94305 }

\begin{abstract}
We study the acoustic phonon response of crystals hosting a gapped time-reversal symmetry breaking
electronic state. The phonon effective action can in general acquire a dissipationless ``Hall'' viscosity, which
is determined by the adiabatic Berry curvature of the electron wave function. This Hall viscosity endows the system with
a characteristic frequency, $\omega_v$; for acoustic phonons of frequency $\omega$, it shifts the phonon spectrum
by an amount of order $(\omega/\omega_v)^2$ and it mixes the longitudinal and transverse acoustic phonons with a
relative amplitude ratio of $\omega/\omega_v$ and with a phase shift of $\pm \pi/2$, to lowest order in $\omega/\omega_v$.
We study several examples, including the integer quantum Hall states, the quantum anomalous Hall state in
Hg$_{1-y}$Mn$_{y}$Te quantum wells, and a mean-field model for $p_x + i p_y$ superconductors.
We discuss situations in which the acoustic phonon response is directly related to the gravitational response, for which
striking predictions have been made. When the electron-phonon system is viewed as a whole, this provides an
example where measurements of Goldstone modes may serve as a probe of adiabatic curvature
of the wave function of the gapped sector of a system.
\end{abstract}
\maketitle

\section{Introduction}

One of the most important discoveries in condensed matter physics has been that there are distinct states of matter
that are distinguished not by their patterns of symmetry breaking, but by their topological order.\cite{wen04} Such topological states of matter (TSM)
cannot be described by local order parameters, but can sometimes be characterized by quantized topological responses to external fields.
For example, the quantum Hall states\cite{klitzing1980,tsui1982}---the first topological states
discovered in nature---can be characterized by their quantized Hall conductance.
The three-dimensional time-reversal invariant topological insulators discovered more
recently\cite{fu2007b,moore2007,roy2009} can be characterized by a topological magneto-electric effect\cite{qi2008b,EM0905}.
More generically, topological insulators in arbitrary dimensions can be characterized by topological responses to
electromagnetic fields\cite{qi2008b}. However, many TSM cannot be characterized by electromagnetic response.
For example in topological superconductors the charge conservation symmetry is effectively broken, and the
electromagnetic field is screened. Thus more generic response properties need to be investigated
in order to distinguish different topological states.\cite{RM1036,NR1154,WQ1127}

In this paper, we propose a response property, the ``phonon Hall viscosity," for gapped states that break
time-reversal symmetry. For a quantum liquid, the viscosity tensor $\eta_{ijkl}$ is defined by the linear
response $T_{ij}=\eta_{ijkl}v_{kl}$, with $T_{ij}$ the stress tensor, and $v_{ij}=\frac12 (\partial_iv_j+\partial_jv_i)$ the
gradient of the velocity field $v_{i}$. Usually, a finite viscosity indicates dissipation in the system, similar
to a finite resistivity. However the viscosity can have a dissipationless component, associated with the
part of $\eta_{ijkl}$ that is anti-symmetric under exchange of the first and second pair of indices.\cite{PL1981} 
This Hall viscosity can only exist in a system that breaks time-reversal symmetry 
and is analogous to the dissipationless Hall resistivity.
The Hall viscosity has appeared in the hydrodynamic theory of the A-phase of He-3\cite{Volovik1984}
and was studied for quantum Hall liquids by J. E. Avron  {\it et. al.}\cite{Avron95,Avron98}. It has 
since been studied for various $(2+1)$-d topological states including IQH states\cite{Avron95}, the $(2+1)$-d Dirac
model\cite{Hughes11}, fractional quantum Hall (FQH) states\cite{Read09,Haldane09,ReadRezayi10,HS0951} and
$p_x+ip_y$-paired topological superconductors\cite{Read09}. 
 %The QH states are considered as incompressible liquid of electrons
The Hall viscosity provides a probe of gapped time-reversal symmetry breaking states in the charge neutral channel, which does not require charge conservation and thus may be a suitable response for TSC and more generic
TSM that cannot be characterized by topological electromagnetic response. In particular, it was recently
proposed that the Hall viscosity of a rotationally invariant system is related to the angular momentum
carried by each quasi-particle of the system\cite{Read09,Son1}, which is in turn proportional to the
``topological shift" of the topological field theory on the sphere\cite{WenZeeShift92}.  However, the
Hall viscosity is only defined for a liquid in continuum, since the stress tensor is a momentum current
which is ill-defined if continuous translation symmetry is broken by the lattice. The discussions of Hall viscosity in
the literature have been treating the electron system as a liquid without explicitly considering the lattice effects. This
approximation is in general problematic since the Hall viscosity intrinsically depends on a length scale,
and there are two natural length scales in a gapped system: a length scale associated with the energy gap, and
a different length scale associated with the electron density. In general, the Hall viscosity will depend on
both of these length scales and will therefore depend on non-universal short range physics.
Even if the results in the continuum approximation are applicable in
some cases, we are facing another problem of how to observe the Hall viscosity in general since it is difficult to measure
the velocity and stress of the electron liquid. To distinguish the Hall viscosity defined in this traditional way with the phonon Hall viscosity that we study in this work, we refer to the Hall viscosity of the continuum electron liquid as the
gravitational Hall viscosity since the viscosity tensor of the electron liquid can be considered as a response to an external deformation of the spatial metric $g_{ij}$\cite{Avron95}.

To solve these problems of the gravitational Hall viscosity, we alternatively define the phonon Hall viscosity, which
is the adiabatic response of the electron state to the deformation of the crystal, {\it i.e.}, to acoustic phonons. Instead of
the stress tensor which couples to the deformation of the spatial metric and is only well-defined in the continuum,
the deformation of the crystal and the electron-phonon coupling are always well-defined.
%defining the minimal coupling to metric which is only well-defined in the continuum limit, we can
%consider the coupling of the electron system to the nuclei which is always well-defined.
The linear response of the electron liquid to the crystal deformation leads to a correction to
phonon dynamics. The strain field $u_{ij}=\frac12(\partial_i u_j+\partial_j u_i)$ plays the role of the spatial
metric $g_{ij}$, with $u_i$ the displacement field of the nuclei. The phonon Hall viscosity is defined
by the linear response $\left\langle \frac{\partial H}{\partial
    u_{ij}}\right\rangle=\lambda_{ijkl} u_{kl} + \eta_{ijkl}\dot{u}_{kl}$ with
$H$ the electron Hamiltonian depending on the lattice strain $u_{ij}$, $\lambda_{ijkl}$ the elastic moduli, and $\eta_{ijkl}$ the
phonon Hall viscosity. The phonon Hall viscosity is well-defined for all gapped states with a regular lattice background, 
and is physically observable through phonon dynamics.

%The dissipationless phonon Hall viscosity can appear in generic gapped time-reversal symmetry breaking
%states. While it also depends on non-universal physics of the material under study, its feature is that
%it is a direct probe into the adiabatic Berry curvature of the electronic wave functions. As such, it
%may be prove useful as a novel probe into quantum many-body states and their possible topological behavior.

In the rest of this paper, we will first give a general definition of the phonon Hall viscosity in Sec. II and show
how it relates to the adiabatic Berry curvature of the many-body electron wave function. In Sec. III we
study the phonon Hall viscosity in several example systems such as quantum Hall states, quantum anomalous Hall
states and $(p+ip)$ superconductors. We make comparisons between the phonon Hall
viscosity and the gravitational Hall viscosity in the continuum limit, and in certain examples, we analyze
features of the Hall viscosity that are universal or depend only on the low-energy theory of the electronic system.
Finally in Sec. IV we will discuss the physical consequences of the phonon Hall viscosity, give order of magnitude
numerical estimates, and discuss experimental prospects for the observation of phonon Hall viscosity.

\section{Definition of the Phonon Hall Viscosity}

\subsection{Effective action of acoustic phonons}
\label{effActionSec}

In this section we will define the phonon Hall viscosity based on a generic discussion of acoustic phonon dynamics.
The dynamics of acoustic phonons can be described using the long-wavelength effective action in terms of the displacement
fields $\v{u}(\v{r})$, which describe the displacement of an atom from its original location. For an
insulator, dissipation can be ignored at frequencies below the energy gap and the long-wavelength elastic response is described by
an effective action for the displacement fields $\v{u}$, which can be obtained by integrating out the electrons:
$e^{-S_{eff} [\v{u}]} = \int \mathcal{D}c^\dagger \mathcal{D}c e^{-S[\v{u},c,c^\dagger]}$. When the electrons form a gapped state,
the phonon effective action is, to lowest order,
\begin{align}
\label{phAction}
S_{eff} = \frac{1}{2}\int d^d x dt (\rho \partial_t u_j \partial_t u_j -  \lambda_{ijkl} \partial_i u_j \partial_k u_l ),
\end{align}
where $\lambda_{ijkl} = \lambda_{klij} = \lambda_{jikl}$ are the elastic moduli; its symmetry under interchange of the first or second pair of
indices follows from invariance of the energy under rigid rotations. For a gapless state, such as a metal or a magnet,
the phonon effective action will not be a local theory in terms of the strain fields. In the presence of time-reversal
symmetry breaking, ``Hall'' viscosity terms are allowed:
\begin{align}
\label{phHallAction}
\delta S_{H} = \frac12\int d^d x dt \eta_{ijkl} \partial_i u_j \partial_k\dot{u}_l,
\end{align}
where $\eta_{ijkl} = -\eta_{klij}$. For an inversion symmetric system, this is the only additional term that can
be added, to cubic order in momenta. Anharmonic corrections to the acoustic phonon dynamics go like $(\partial u)^4$, 
so they are $\mathcal{O}(k^4)$, implying that the Hall viscosity term may be distinctly measurable since it is lower order
in momenta. Impurities in the crystal may be difficult to treat, but as we discuss in the Section \ref{expDisc}, their contributions are not sensitive
to the sign of the time-reversal symmetry breaking of the electronic system and may be separated from Hall viscosity contributions. 

Up to total derivatives, (\ref{phHallAction}) only depends on $\eta_{ijkl} + \eta_{kjil}$:
\begin{align}
\delta S_H = \frac12\int d^d x dt [\eta_{ijkl}^{+} \partial_i u_j \partial_k\dot{u}_l + \eta_{ijkl}^- \partial_i (u_j \partial_k \dot{u}_l)],
\end{align}
where $\eta_{ijkl}^{\pm} = \frac12 (\eta_{ijkl}\pm \eta_{kjil})$.
The boundary terms may have interesting consequences for the surface waves of a medium with such a Hall viscosity, but we will
ignore them in this paper. In some cases it will be conceptually more clear to write the above action
in terms of the strain tensor $u_{ij} \equiv \frac12 (\partial_i u_j + \partial_j u_i)$
and the rotation tensor $m_{ij} \equiv \frac12 (\partial_i u_j - \partial_j u_i)$:
\begin{align}
\label{phHallAction2}
\delta S_{H} = \frac12\int d^d x dt [&\eta_{ijkl}^{SS} u_{ij} \dot{u}_{kl} + \eta_{ijkl}^{AA} m_{ij} \dot{m}_{kl}
\nonumber \\
& + 2 \eta_{ijkl}^{SA} u_{ij} \dot{m}_{kl} ],
\end{align}
where $\eta_{ijkl}^{SS} = \eta_{jikl}^{SS} = -\eta_{klij}^{SS}$, $\eta_{ijkl}^{AA} = -\eta_{jikl}^{AA} = -\eta^{AA}_{klij}$, and
$\eta^{SA}_{ijkl} = \eta_{jikl}^{SA} = -\eta_{ijlk}^{SA}$ can all be deduced from $\eta_{ijkl}$.
%Note that it is only $\eta_{ijkl}^{SS}$ that has the same symmetry properties under interchange of indices as $\eta_{ijkl}^{gr}$.
For an isotropic three-dimensional system, $\eta_{ijkl}$ must vanish. In two dimensions, $\eta^{AA}$ always vanishes;
for an isotropic 2D system, or one with $\pi/4$ rotation symmetry, $\eta^{SS}$ and $\eta^{SA}$ each reduce to a
single number: $\eta^H \equiv \eta_{xxxy}^{SS} = \eta_{xyyy}^{SS}$ \cite{Avron95,Avron98}
and $\eta^M \equiv \eta_{xxxy}^{SA} = \eta_{yyxy}^{SA}$. In this case,
\begin{align}
\label{isoSH}
\delta S_H =  2\int d^2 x dt &[\eta^H (u_{xx} - u_{yy}) \dot{u}_{xy}\nonumber\\
&+ \eta^M (u_{xx} + u_{yy}) \dot{m}_{xy}],
\end{align}
where $\eta^H = \frac12(\eta_{xxxy} + \eta_{xxyx}) = \frac12(\eta_{xyyy} + \eta_{yxyy})$
and $\eta^M = \frac12(\eta_{xxxy} - \eta_{xxyx})= \frac12(\eta_{xyyy} - \eta_{yxyy})$.
It also follows that, up to boundary terms, ($\ref{isoSH})$ is equivalent to:
\begin{align}
\label{phHallActionIso}
\delta S_H =  \int d^2 x dt &[\eta_{xxxy} (u_{xx} - u_{yy}) \dot{u}_{xy}\nonumber\\
&+ \eta_{xxxy} (u_{xx} + u_{yy}) \dot{m}_{xy}].
\end{align}
To obtain this, we have used the fact that, up to boundary terms, $\eta_{xxyx}$ does
not contribute to the action.

%({\bf I feel the following two sentences are contradictory with Eq. 7 which says $\eta^H=\eta^M$. Maybe remove the two sentences.})
%For systems where the Hamiltonian is explicitly invariant under rotations, such as for clean quantum Hall states, $\eta^M = 0$.
%For systems that either explicitly or spontaneously break rotational symmetry, the electronic Hamilonian may depend on $m_{xy}$,
%so one obtains $\eta^M \neq 0$.

%The above dependence on $m_{ij}$ may occur even in a rotationally
%invariant system, since it is a Berry phase effect and does not affect the energy of the system.
%While this term is irrelevant in the RG sense for the above
%effective action, it is still important.
%As we will discuss in more detail below,
%the physical consequence of this phonon Hall viscosity is to mix the transverse and longitudinal
%sound modes. The Hall viscosity endows the system with a characteristic frequency scale, $\omega_{v} = \rho c^2/\eta$,
%where  $\rho$ is the mass density, and $c$ is the speed of sound. The amount of mixing between the longitudinal and
%transverse modes at a frequency $\omega$ is then determined by the dimensionless ratio $\omega/\omega_v$.
%To lowest order in $\omega/\omega_v$, there is a phase shift of $\frac{\pi}{2}$  between these modes.

\subsection{Phonon Hall viscosity as a response property of the electron system}

In the adiabatic approximation, it is assumed that the motion of the lattice is infinitely slow compared with the motion of the
electrons, so that at any moment, the electrons are in their ground state with respect to that particular instantaneous configuration
of the lattice. Within this approximation, the effect of lattice displacements is to alter the parameters
in the effective tight-binding model for the electron system. The dependence of these parameters on the atom positions can be calculated using standard
\it ab initio \rm methods. Thus, in the adiabatic approximation, the electrons will be described by an effective tight-binding
Hamiltonian, where the lattice displacements appear as external parameters: $H_{t.b.}[\{\v{u}_i \}]$. $\v{u}_n$ is the displacement
of the $n$th atom from its original location. We may also view $H_{t.b.}$ as a function of the Fourier components
$\v{u}_{\v{q}} = \frac{1}{\sqrt{N_{site}}}\sum_{\v{n}} \v{u}_{\v{n}} e^{i \v{q} \cdot \v{n}}$. Then, the following linear response formula
\begin{align}
\eta_{ab}(\v{q},\omega) = \frac{1}{\omega}
\frac{1}{L^d} \int dt e^{i \omega t} \left\langle \left[\frac{\partial H_{t.b.}}{\partial u_{\v{q},a}}(t), \frac{\partial H_{t.b.}}{\partial u_{-\v{q},b}}(0)\right] \right\rangle
\end{align}
gives an additional term to the acoustic phonon effective action of the form
\begin{align}
\label{fullPhResp}
\delta S = \frac12 \int d^{d+1}x d^{d+1}x' \eta_{ab}(x - x') u_{a}(x) \dot{u}_{b}(x'),
\end{align}
where $x$ here is a $(d+1)$-component vector including space and time,
$\eta_{ab}(x)$ is the Fourier transform into real space-time of $\eta_{ab}(\v{q}, \omega)$,
and we have taken the continuum limit to get the displacement field $\v{u}(x)$.
The leading order term that is independent of uniform displacements
$\v{u} \rightarrow \v{u} + \v{a}$ is given by (\ref{phHallAction}). Starting from
(\ref{fullPhResp}), we find
%Expanding $\eta_{ab}(\v{q},\omega)$ in powers of $\v{q}$ and converting to real-space, the effective action is
%\begin{align}
%\delta S = \frac12 \int d^d x dt [&\eta_{ij} u_i \dot{u}_j + \eta_{ijk} \partial_i u_j \dot{u}_k
%\nonumber \\
%&+ \eta_{ijkl} \partial_i u_j \partial_k u_l + \cdots].
%\end{align}
%where $\eta_{ij} = \eta_{ij}(q=0)$, $i\eta_{ijk} = (\partial \eta_{jk} /\partial q_i)|_{q = 0}$,
%$\eta_{ijkl} = \frac{1}{2} (\partial^2 \eta_{kl}/\partial q_i \partial q_k)_{q=0}$.
%Since the lattice tight-binding Hamiltonians are all invariant under a uniform displacement $\v{u}_n \rightarrow \v{u}_n + \v{a}$ of
%all the atoms, it explicitly only depends on the distortion tensor
%$\partial_i u_j$ and its derivatives. As a result, the first and second terms above will vanish. Such terms may
%appear, however, when the positive charge of the ions is considered; we discuss them elsewhere. ({\bf refer to the section where this is discussed.}) Thus the Hall viscosity is given by
\begin{align}
\eta_{ijkl} = \frac{1}{2} \lim_{\omega \rightarrow 0} \lim_{q \rightarrow 0} \frac{\partial}{\partial q_i} \frac{\partial}{\partial q_k} \eta_{jl} (\v{q}, \omega)
\end{align}
For spatially homogeneous deformations, the distortion tensor $w_{ij} \equiv \partial_i u_j$ is a constant.
To calculate $\eta_{ijkl}$, it will be more convenient to take $w_{ij}$ to be a constant and to treat it as a parameter in $H_{t.b.}$. Then,
\begin{align}
\label{phHallVisc}
\eta_{ijkl} = &\frac{1}{2} \lim_{\omega \rightarrow 0} \frac{1}{\omega}
\frac{1}{L^d} \int dt e^{i \omega t} \left\langle \left[\frac{\partial H_{t.b.}}{\partial w_{ij}}(t), \frac{\partial H_{t.b.}}{\partial w_{kl}}(0)\right] \right\rangle
\nonumber \\
& + (i \leftrightarrow k)
\end{align}
For spatially inhomogenous deformations, we can continue to use the DC response (\ref{phHallVisc}) 
instead of the exact AC response, as long as the acoustic phonon frequency is much less than the electronic energy gap.

It is well-known that the adiabatic response of a Hamiltonian to changes in some parameter is directly related
to Berry curvature \cite{WilczekGP} of the ground state wave function. For a Hamiltonian
$H[\{\lambda_i\}]$ that depends on a set of parameters $\{\lambda_i\}$, we have
\begin{align}
\left\langle \frac{\partial H}{\partial \lambda_i} \right\rangle = \frac{\partial E}{\partial \lambda_i} + \Omega_{ij} \dot{\lambda_j},
\end{align}
where $\Omega_{ij}$ is the Berry curvature of the ground state wave function. Thus
$\eta_{ijkl}$ is given by the Berry curvature associated with adiabatically varying the distortion tensor $w_{ij} \equiv \partial_i u_j$
as external parameters:
\begin{align}
i\eta_{ijkl} = \frac{1}{2} \left( \frac{\partial}{\partial w_{ij}} \langle \psi | \frac{\partial}{\partial w_{kl}} |\psi \rangle - \frac{\partial}{\partial w_{kl}} \langle \psi | \frac{\partial}{\partial w_{ij}} |\psi \rangle +
(i \leftrightarrow k) \right),
\end{align}
where $|\psi\rangle$ is the ground state of the tight-binding Hamiltonian $H_{t.b.}$.

\section{Examples}

In this section, we will study some examples of systems with a phonon Hall viscosity, including electrons hopping among
the s-orbitals of a square lattice in a background magnetic field, a simple model for the quantum anomalous Hall state in
HgMnTe quantum wells, and a simple mean-field model of a spinless $p_x +ip_y$ superconductor. In certain limits
we compare the phonon Hall viscosity of these systems with their conventional Hall viscosity studied in the literature.
% As we discuss, in some situations the phonon Hall viscosity is directly related to the gravitational Hall viscosity.

\subsection{Hofstadter Model}

%In some cases, the phonon Hall viscosity $\eta_{ijkl}$ is directly
%related to the long-wavelength gravitational response $\eta_{ijkl}^{gr}$ of the electron system. As an example,
Consider a square lattice with nearest and next-nearest neighbor hopping:
\begin{align}
H = -\frac{1}{2} \sum_{\langle ij \rangle} t_{ij} e^{i A_{ij}} c_i^\dagger c_j - \frac{1}{2} \sum_{\langle \langle ij \rangle \rangle} \tilde{t}_{ij}e^{iA_{ij}} c_i^\dagger c_j + h.c.
\end{align}
Consider hopping among s-wave orbitals, in which case $t_{ij}$ and $\tilde{t}_{ij}$ depend only on
the distance $\left|{\bf r}_j-{\bf r}_i\right|$ between atoms. To leading order in the crystal deformations,
$t_{i,i+\hat{\bf x}}\simeq t+t'u_{xx}$, $t_{i,i+\hat{\bf y}}\simeq t+t'u_{yy}$, and
$\tilde{t}_{i,i+\hat{\bf x} \pm \hat{\bf y}} =\tilde{t} + \tilde{t}' ( \frac12 (u_{xx} + u_{yy}) \pm u_{xy})$.
If $t(r)$ is the hopping matrix element between the s-wave orbitals that are a distance $r$ apart,
$t \equiv t(a)$, $\tilde{t} = t(\sqrt{2} a)$, $t' = a \frac{\partial t}{\partial r}|_{a}$, $\tilde{t}' \equiv \sqrt{2} a \frac{\partial t}{\partial r}|_{\sqrt{2} a}$.
%({\bf try to explain more explicitly the form of $\tilde{t}_{ij}$.}) $t'$ and $\tilde{t}'$ are their spatial derivatives.
In the absence of a background electromagnetic field and for constant lattice displacements,
the Hamiltonian can be written in momentum space as $H = \sum_{\bf k} \epsilon_{\bf k} c_{\bf k}^\dagger c_{\bf k}$.
We note that a significant effect of the strain fields on the energy of the system is to change the on-site energy of atomic orbitals.
However, this contribution does not affect the Hall viscosity below, so we ignore it. 

In the continuum limit, the dispersion is, up to a constant, %can be expanded to first order in the strain tensor. In the continuum limit,
\begin{align}
\label{kinEn}
\epsilon_k \simeq \frac{1}{2m^*}k_i k_j g_{ij} - (t' + \tilde{t}') (u_{xx} + u_{yy}),
\end{align}
where the effective mass is defined by $\frac{1}{2m^*} \equiv (t/2 + \tilde{t})$, and
$g_{ij} =  \delta_{ij} + \delta g_{ij}$,
\begin{align}
\label{kinMetric}
\delta g &= 2 m^*\frac{\tilde{t}'}{2}\left(
\begin{matrix}
(1 + \frac{t'}{\tilde{t}'})  u_{xx} + u_{yy} &  2 u_{xy} \\
2 u_{xy} & (1 + \frac{t'}{\tilde{t}'}) u_{yy} + u_{xx} \\
\end{matrix} \right)
\end{align}
In the presence of a gauge field, we take $k \rightarrow -i D \equiv -i (\partial - i A)$,
so the effective theory becomes
\begin{align}
H = - \frac{1}{2m^*} g_{ij} D_i D_j -  (t' + \tilde{t}') (u_{xx} + u_{yy}).
\end{align}
The phonon Hall viscosity is then related directly to the gravitational Hall viscosity of the
electronic fluid\cite{Avron95}. We find
\begin{align}
\label{HallViscRel}
\eta^{H} \equiv \eta_{xxxy}^{SS} = (t/2 + \tilde{t})^{-2} \frac{\tilde{t}' t'}{4} \eta^{H}_{gr}.
\end{align}
For $N_L$ filled Landau levels, $\eta^{H}_{gr} = N_L\hbar n/4$, where $n$ is the density of electrons,
is the Hall viscosity of the electron liquid when the crystal is ignored.\cite{Read09,Avron95}
%{\bf with $\eta^H_{gr}=??$} the Hall viscosity of electron liquid while the crystal is ignored.
The prefactor  $(t/2 + \tilde{t})^{-2}\tilde{t}' t'$ can explicitly be verified
to be of order one for typical s-wave orbitals and typical separations between atoms; it can also
be calculated  fairly precisely using \it ab initio \rm methods. Note that since the Hamiltonian depends
only on the strain field $u_{ij}$, the phonon effective action also only depends on $u_{ij}$ and there
is no dependence on the rotation tensor $m_{ij}$; i.e. $\eta^{SA} = \eta^{AA} = 0$.

Therefore, we see that the result of \Ref{Avron95, Levay95} for integer quantum Hall (IQH) states
has a direct effect in the phonon response, which, as will be discussed in Sec. \ref{sec:observation}, is a directly measurable
physical quantity. For higher density systems, we cannot take the continuum limit;
the phonon Hall viscosity is still a well-defined quantity that is calculable
through linear response theory, but the previously defined ``gravitational Hall viscosity"of \Ref{Avron95,Levay95} is not well-defined.

\subsection{Quantized anomalous Hall state and Hg$_{1-y}$Mn$_y$Te quantum wells}

Here we will calculate the phonon Hall viscosity for Hg$_{1-y}$Mn$_y$Te quantum wells, which
exhibit a quantized anomalous Hall (QAH) state for certain thicknesses of the quantum well and
spin polarization of the Mn ions \cite{LQ0802}. The QAH state is a band insulator
that exhibits a quantized Hall conductance in the absence of a net magnetic field. The first lattice
model for such a state was introduced by Haldane\cite{haldane1988}, and since then it has been proposed
to be realized in  Hg$_{1-y}$Mn$_y$Te quantum wells.\cite{qi2005,liu2009C,yu2010} As the quantum well thickness and
the magnetization of the Mn ions is tuned, the system can be tuned between different topological states:
a quantum spin Hall state, QAH states, and the topologically trivial state.

At the topological phase transitions, the phonon Hall viscosity exhibits non-analyticities that can be
accounted for in the continuum Dirac approximation. In what follows, we will calculate the phonon Hall viscosity
for physically realistic parameters, we will isolate the universal contributions that
depend only on the low energy physics near the Dirac cones and we make contact with the calculations
of \Ref{Hughes11} for the regularized gravitational Hall viscosity of the continuum Dirac model.

The model for Hg$_{1-y}$Mn$_y$Te quantum wells is given by a four-band Bloch Hamiltonian:
\begin{align}
H(\v{k}) &=\left(
\begin{matrix}
h_+(\v{k}) & 0 \\
0 & h_-(\v{k}) \\
\end{matrix} \right),
\end{align}
where the two-band Bloch Hamiltonians can be expanded in terms of Pauli matrices
$h_{\pm}(\v{k}) = \epsilon_\pm(\v{k}) \mathbb{I} + \v{d}_\pm(\v{k}) \cdot \v{\sigma}$,
and $h_-(\v{k}) = h_+^*(- \v{k})$. In the continuum limit and in the absence of lattice distortions,
expanding near the $\Gamma$ point $\v{k} = (0,0)$, we have
\begin{align}
d_{\pm,x} + id_{\pm, y} &= A (\pm k_x + i k_y),
\nonumber \\
d_{\pm, z} &= M_{\pm} - B (k_x^2 + k_y^2),
\nonumber \\
\epsilon(k) &= C_{\pm} - D(k_x ^2 + k_y^2),
\end{align}
where $M_{\pm} = M \pm \delta M/2$.
The parameters $A$, $B$, $C$, $D$, $M$, and the lattice spacing $a$ are given
in \Ref{BHZ06, Koenig08} for HgCdTe/HgTe
quantum wells and the relevant ones are listed in Table \ref{QAHpar}. $\delta M$
%$G_E$ and $G_H$ also
depends on Mn doping and spin polarization, as discussed in \Ref{LQ0802}.

The phonon Hall viscosity will be a sum of the contributions of each of the two blocks:
\begin{align}
\label{hallViscSum}
\eta_{ijkl} = \eta_{ijkl}^{+} + \eta_{ijkl}^-,
\end{align}
where
\begin{align}
\label{kuboD}
\eta_{ijkl}^\pm = &\frac12 \frac{\hbar}{8 \pi^2} \frac{1}{a^2} \int d^2k \hat{\bf d}_{\pm} \cdot
\left(\frac{\partial \hat{\bf d}_\pm}{\partial (\partial_i u_j)} \times \frac{\partial \hat{\bf d}_\pm}{\partial (\partial_k u_l)}\right)
\nonumber \\
&+ (i \leftrightarrow k)
\end{align}

In order to calculate $\eta_{ijkl}$, we need to obtain the Hamiltonian as a function of lattice distortions.
To do this, observe that the blocks $h_{\pm}(\v{k})$ are composed of the spin-orbit coupled states
$|s,\pm\frac12 \rangle$ and $|p_x \pm ip_y ; \pm\frac12\rangle$.\cite{BHZ06}
Concentrating on a single $2 \times 2$ block -- for definiteness consider $h_+$ -- the Hamiltonian
is written as:
\begin{align}
H_+ = \frac12 &\sum_{n,i} c_n^\dagger (\tilde{t}_i \mathbb{I}  + t_i \sigma^z + \v{e}_i \cdot \v{\sigma}) c_{n+\hat{i}}
\nonumber \\
&+ m_+ \sum_n c_n^\dagger \sigma^z c_n + h.c.,
\end{align}
where $i = x, y$ and $n$ labels the sites of a two-dimensional square lattice. $\v{\sigma}$ is the vector
of Pauli matrices and the hopping parameters are, to first order in lattice distortions,
\begin{align}
\tilde{t}_i &= \tilde{t} + a\tilde{t}' \partial_i u_i,
\nonumber \\
t_i &= t + at' \partial_i u_i,
\nonumber \\
\v{e}_x &= i (\lambda + a\lambda' \partial_x u_x) (\hat{x} + \partial_x u_y \hat{y})
\nonumber \\
\v{e}_y &= i (\lambda + a\lambda' \partial_y u_y) (\partial_y u_x \hat{x} + \hat{y})
\end{align}
%\begin{widetext}
%\begin{align}
%H_+ = &\frac12 \sum_{n,a} [c_n^\dagger (t + t' \partial_a u_a) \sigma^z c_{n+\hat{a}} +  c_n^\dagger  (t_2 + t'_2 \partial_a u_a) c_n + h.c.]
%+ m_+ \sum_n c_n^\dagger \sigma^z c_n
%\nonumber \\
%&+ \frac{i}{2} \sum_{n}  [(\lambda + \lambda' \partial_x u_x)  c_n^\dagger ( \sigma_x +  \partial_x u_y \sigma_y) c_{n+ \hat{x}}
%+  (\lambda + \lambda' \partial_y u_y)  c_n^\dagger ( \sigma_y +  \partial_y u_x \sigma_x) c_{n+ \hat{y}}] + h.c.
%\end{align}
%\end{widetext}
%In momentum space, this is written as
%\begin{align}
%H = \sum_k c_k^\dagger [\epsilon(\v{k}) \mathbb{I} + \v{d}(\v{k}) \cdot \v{\sigma} ] c_k,
%\end{align}
%where
%\begin{align}
%\epsilon(k) &= (t_2 + t_2' u_{xx} ) \cos k_x + (t_2 + t_2' u_{yy} ) \cos k_y
%\nonumber \\
%d_x &= (\lambda + \lambda' u_{xx}) \sin k_x + \lambda \partial_y u_x \sin k_y,
%\nonumber \\
%d_y &= (\lambda + \lambda' u_{yy}) \sin k_y + \lambda \partial_x u_y \sin k_x,
%\nonumber \\
%d_z &= (t + t' \partial_x u_x ) \cos k_x + (t + t' \partial_y u_y) \cos k_y + (m - 2t),
%\end{align}
The lattice parameters $\lambda$, $t$, $\tilde{t}$, and $m$ used above are related to the continuum parameters $A$, $B$, $C$, $D$, and $M$
through: $\lambda = A/a$, $M_{\pm} = m_{\pm} + 2t$, $B = a^2 t/2$, $C = 2 \tilde{t}$, and $D = \tilde{t}/2$.
The hopping parameters are functions of the distance between neighboring atoms; the prime indicates
a derivative with respect to this distance. The upper $2\times 2$ block has topological phase transitions
as $\frac{a^2 M_+}{2B}$ is tuned. The Chern number $C_1$ of the lower of the two bands is:
\begin{align}
C_1 = \left\{ \begin{array}{lll}
1 & \text{ for } & 2 < \frac{a^2 M_+}{2B} < 4  \\
-1 & \text{ for } & 0 < \frac{a^2 M_+}{2B} < 2 \\
0 & & \text{ otherwise }
\end{array} \right.
\end{align}
%$\frac{a^2 M_+}{2B} = 2, 0, 4$; its first Chern number is +1 for $2 < \frac{a^2 M_+}{2B} < 4$, -1 for
%$0 < \frac{a^2 M_+}{2B} < 2$, and 0 otherwise,\cite{qi2008}
and similarly for the lower $2\times 2$ block.
%This model has topological phase transitions at $m = \pm 2 t, 0$;
%ts first Chern number is +1 for $0 < m/t < 2$, -1 for $-2 < m/t < 0$, and 0 otherwise.\cite{qi2008}

From the Kubo formula (\ref{kuboD}), we see that the only non-zero
terms are $\eta_{xxxy}^+ = -\eta_{yyyx}^+$. The full effective action is given by (\ref{phHallActionIso}); making the physically reasonable approximation
%\begin{align}
%\eta_{xxxy} = \frac{\hbar}{8\pi^2} \frac{1}{a^2} \int d^2k \frac{1}{|d|^3} [\lambda \lambda' \sin^2 k_x d_z - \lambda^2 t' \sin^2k_x \cos k_x ]
%\end{align}
$a \lambda' \approx \lambda$ and $a t' \approx t$, we find %({\bf use physical parameters in this equation and Eq. 28})
\begin{align}
%\eta_{xxxy}^+ = \frac{\hbar}{8 \pi a^2} f\left( \frac{\lambda}{t}, \frac{m_+}{t} \right),
\eta_{xxxy}^+ = \frac{\hbar}{8 \pi a^2} f\left( \frac{aA}{2B}, \frac{a^2 M_+}{2B} \right),
\end{align}
where $f$ is a function of dimensionless parameters:
\begin{widetext}
\begin{align}
\label{fIntegral}
f(\alpha,\beta) = \int \frac{\alpha^2 \sin^2 k_x (\beta -2 + \cos k_y) d^2k }{[\alpha^2(\sin^2 k_x + \sin^2 k_y) + (\beta - 2 + (\cos k_x + \cos k_y))^2]^{3/2}}.
\end{align}
\end{widetext}

\begin{figure}[b]
\centerline{
\includegraphics[width=3.2in,height=2.3in]{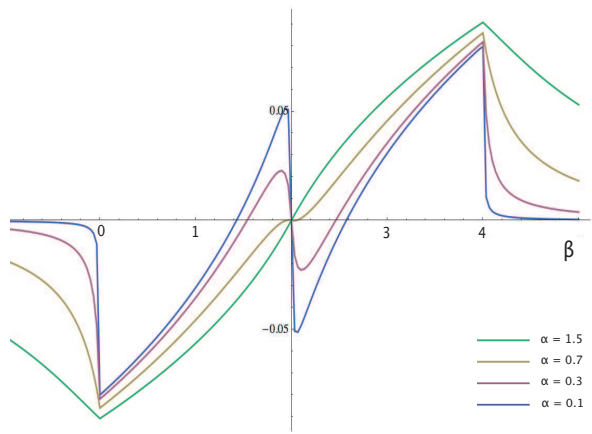}
}
\caption{The function $\frac{1}{8\pi^2} f(\alpha, \beta)$, plotted as a function of $\beta$ for various choices of $\alpha$.
\label{QAH1}
}
\end{figure}

In Fig. \ref{QAH1}, we plot the function $\frac{1}{8 \pi^2} f(\alpha , \beta)$ as a function of $\beta$
for various choices of $\alpha$. Note $f$ has non-analyticities at the quantum phase transitions
$\beta = 0, 2$, and $4$.
%The low energy physics is determined by the parameters of the continuum model, $A$, $B$, $C$, $D$, and $M$, which are
%related to the microscopic parameters by various numerical factors and the lattice spacing.
%As shown in Fig. ??, we see that near the topological phase transition at $m/t = \pm 2$, the jump in $f(\alpha, \beta)$
%is nearly independent of the lattice spacing $a$, reflecting the fact that the behavior arises from low energy
%physics.

\begin{figure}[b]
\centerline{
\includegraphics[width=3.2in,height=2.3in]{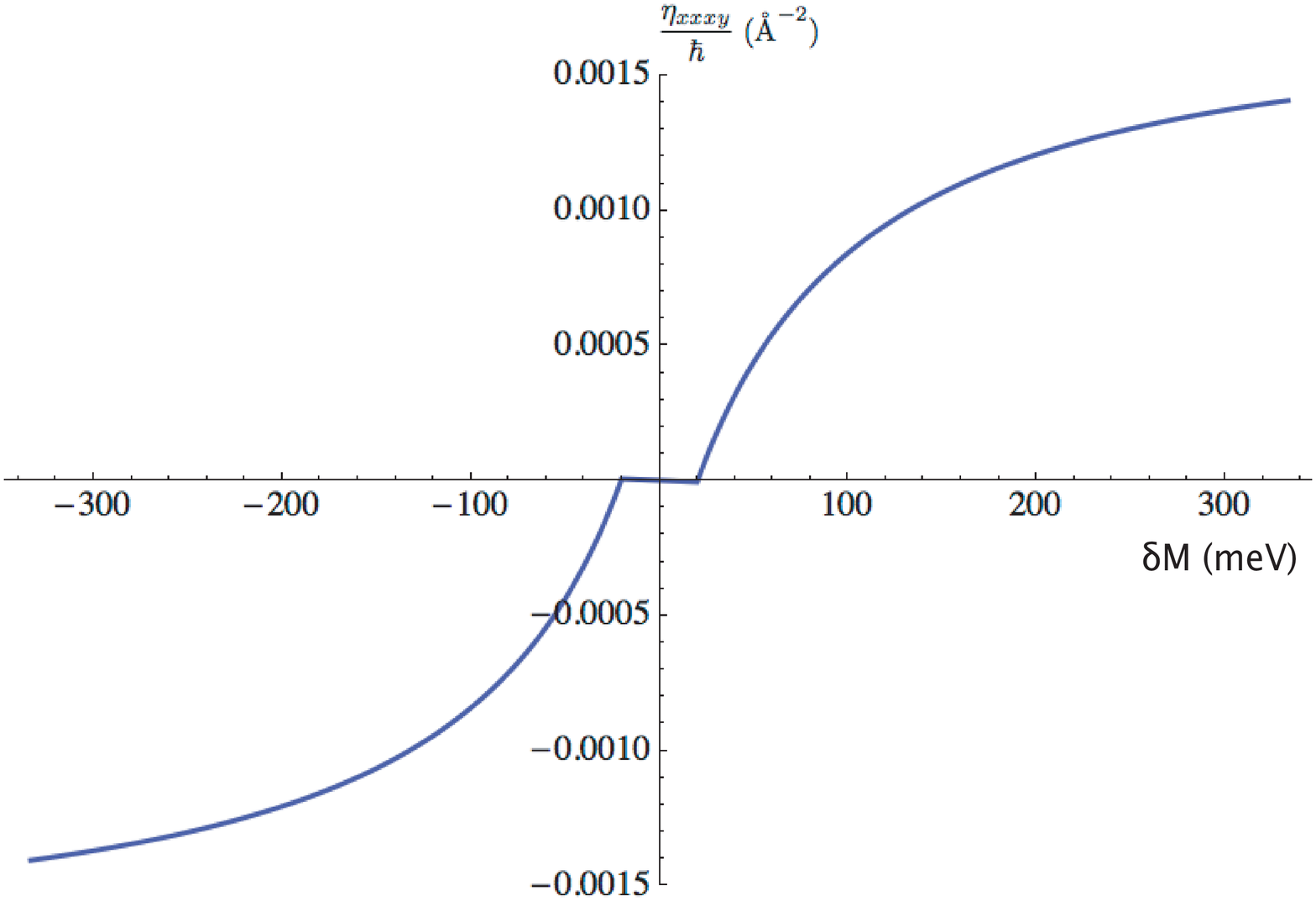}
}
\caption{Plot of $\eta_{xxxy}/\hbar$ as a function of $\delta M$ for fixed $M$ and for realistic parameters
(Table \ref{QAHpar}).
There is a discontinuity in the slope at the transition to the quantum anomalous Hall state, which occurs
at $\delta M = \pm 2 M$.
\label{hallViscA}
}
\end{figure}

\begin{figure}[tb]
\centerline{
\includegraphics[width=3.2in,height=2.3in]{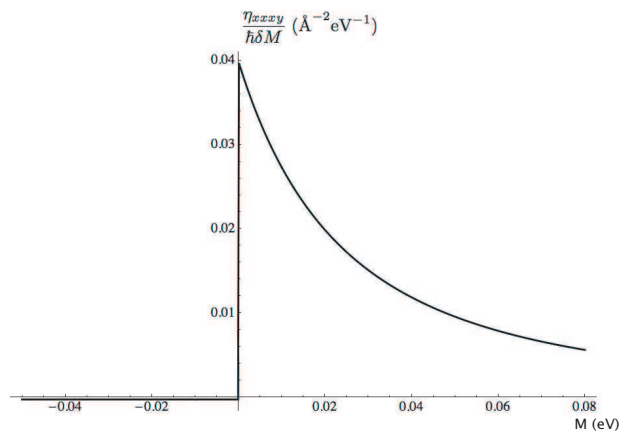}
}
\caption{Plot of $\eta_{xxxy}/\hbar \delta M$ as a function of $M$ for fixed $\delta M = 10^{-7} eV$ and for realistic parameters (Table \ref{QAHpar}). There is a discontinuity at the transition to
the quantum anomalous Hall state, which occurs
at $\delta M = \pm M$.
\label{hallViscC}
}
\end{figure}

The full Hall viscosity is given by the sum of contributions from the two blocks (see \ref{hallViscSum}), which corresponds to taking
the difference of the curve in Fig. \ref{QAH1} for two different values of $m$:
\begin{align}
\eta_{xxxy} = \frac{\hbar}{8 \pi^2} \frac{1}{a^2}  (f_+ - f_-),
\end{align}
where $f_{\pm} = f \left( \frac{Aa}{2B}, \frac{a^2 M_\pm}{2B} \right)$.
We may view $\eta_{xxxy}$ as a function of $M = (m_+ + m_-)/2 - 4B/a^2$ and $\delta M \equiv m_+ - m_-$. In Figs. \ref{hallViscA} - \ref{hallViscC},
we plot $\eta_{xxxy}$ as a function of either $M$ or $\delta M$, using physically realistic parameters
(Table \ref{QAHpar}) and focusing on the region near one of the transitions into the QAH state.
\begin{table}[t]
\begin{tabular}{cccc}
\hline
$A$ (eV $\cdot$ \AA)  & $B$ (eV $\cdot$ \AA$^2$) & $a$ (\AA) & $M$ (eV)  \\
\hline
3.645 & -68.6 & 6.46 & -0.01\\
\hline
\end{tabular}
\caption{Realistic parameters for HgTe quantum wells, taken from \Ref{Koenig08}
\label{QAHpar}
}
\end{table}

In Fig. \ref{hallViscA}, we fix $M$ and plot $\eta_{xxxy}(\delta M)$; experimentally this can be done by tuning an external magnetic field. We observe
a discontinuity in the slope of $\eta_{xxxy}$ as a function of $\delta M$ at the transition into the QAH state. We denote this
discontinuity $\Delta \frac{\partial \eta_{xxxy}}{\partial \delta M}$.
In Fig. \ref{hallViscC}, we fix
%({\bf replace $t$ and $\alpha, \beta$ in this paragraph by physical parameters})
$a^2 \delta M/2B \ll 1$ and plot $\eta_{xxxy} (M)$. This shows a discontinuity $\Delta \eta(M)$.
%As seen in the figures, the Hall viscosity exhibits various non-analyticities at the topological phase transitions.
While a full lattice calculation is required to compute the Hall viscosity, the properties of these
non-analyticities can be accounted for in the continuum Dirac approximation to the above
lattice model. In this approximation, we take $\sin k \approx k$ and $\cos k \approx 1 - k^2/2$ in (\ref{fIntegral}); %({\bf should $k$ be replaced by ${\bf k}a$?})
near the topological phase transition at $\beta = 0$, we have:
\begin{align}
\label{fdirac}
f(\alpha, \epsilon) \approx -\frac{2\pi \alpha^2 |\epsilon| (\epsilon - 1)}{(\alpha^2 - \epsilon)^{2}} + O(\Lambda),
\end{align}
where $\Lambda$ is a high-energy cutoff. While the Hall viscosity will in general depend on $\Lambda$,
the first term above is responsible for the non-analyticities of $f$ and consequently of $\eta_{xxxy}$.
Using (\ref{fdirac}), we can estimate the discontinuity in $\partial \eta_{xxxy}/\partial \delta M$ at the transition,
as shown in Fig. \ref{hallViscA}:
\begin{align}
\left| \Delta  \frac{\partial \eta_{xxxy}}{\partial \delta M} \right| = \frac{\hbar}{\pi} \frac{|B|}{a^2 A^2}.
\end{align}
Similarly, the discontinuity in $\eta_{xxxy}(M)$ for fixed $a^2 \delta M/2B \ll 1$ at the transition is found to be
\begin{align}
|\Delta \eta_{xxxy} |= \left|\frac{\hbar}{\pi} \frac{B}{a^2 A^2} \delta M \right|.
\end{align}

Finally, we note that the discontinuity in the derivative of $\eta_{xxxy}(M)$ can also be computed
using (\ref{fdirac}):
$\Delta \frac{\partial \eta_{xxxy}}{\partial M} \propto [\frac{\partial^2 f}{\partial \epsilon^2}|_{\epsilon = 0^+} - \frac{\partial^2 f}{\partial \epsilon^2}|_{\epsilon = 0^-}]$.
%\begin{align}
%\Delta \frac{\partial \eta_{xxxy}}{\partial M} \approx \frac{\hbar}{8\pi^2} \frac{a^2}{4B^2} \delta M
%[\frac{\partial^2 f}{\partial \epsilon^2}|_{\epsilon = 0^+} - \frac{\partial^2 f}{\partial \epsilon^2}|_{\epsilon = 0^-}.
%\end{align}
We find:
\begin{align}
\Delta \frac{\partial \eta_{xxxy}}{\partial M} \approx \frac{\hbar}{\pi} \delta M \left( \frac{2}{A^2} - \frac{a^2}{4B^2} \right).
\end{align}
The first term above depends only on parameters of the low-energy theory of the state and is independent of
the lattice spacing and other high-energy details. This is essentially the contribution that was found for the
gravitational Hall viscosity in the regularized Dirac model studied in \Ref{Hughes11}.
We see that the other contributions to the Hall viscosity that we find are dependent on the lattice spacing and
other high energy details, which is why they are missed in the regularization of the continuum Dirac model of \Ref{Hughes11}.

\subsection{Interacting states and $p_x + i p_y$ superconductors}

For interacting systems, while there is no universal relationship %in general
between phonon Hall viscosity and the gravitational
Hall viscosity, a major exception occurs in systems with only on-site interactions:
%\begin{align}
$H = \sum_{ij} (t_{ij} e^{iA_{ij}} c_i^\dagger c_j + h.c.) + U \sum_i n_{i \uparrow} n_{i \downarrow}.$
%\end{align}
If the hopping involves s-wave orbitals, then in the dilute limit, where the system can be described
by a continuum interacting theory, the effect of a strain in the lattice is equivalent to a deformation of the gravitational metric.
For such systems, the phonon Hall viscosity is then directly related to the gravitational Hall viscosity,
through a proportionality factor of order unity, as in (\ref{HallViscRel}).
%These models may exhibit a host
%of interesting phases; the continuum limit of such a model with hard core repulsion is often
%taken to be the microscopic starting point for superfluid He-3, which displays spin triplet Cooper pairing
%and topological order.

As an example of an interacting state with a phonon Hall viscosity, we consider a
BCS mean-field description for a $p_x + i p_y$ superconductor, since
such a model may be relevant for the chiral superconductor Sr$_2$RuO$_4$ \cite{MackenzieMaeno03}.
First consider the interaction between the nearest and the next nearest neighbors,
\begin{equation}
U = -V\sum_{\langle ij \rangle} c^\dagger_i c_i c^\dagger_j c_j - \tilde{V} \sum_{\langle\langle ij \rangle\rangle} c^\dagger_i c_i c^\dagger_j c_j.
\end{equation}
Taking only the Cooper channel of the interaction yields:
\begin{align}
U = - \sum_{\bf{k},\bf{k}'} V_{\bf{k} \bf{k}'} c_{\bf{k'}}^\dagger c_{-\bf{k}'}^\dagger c_{-\bf{k}} c_{\bf{k}},
\end{align}
where
\begin{align}
V_{{\bf k}{\bf k}'} = &\frac{1}{N}[2(V+\tilde{V})+(V'+\tilde{V}') (u_{xx} + u_{yy})
\nonumber \\
&- (V/2 + \tilde{V})(k_i-k'_i) (k_j-k'_j) \tilde{g}_{ij}].
\label{EQ:UMetric}
\end{align}
$\tilde{g}_{ij} = \delta_{ij} + \frac{\tilde{V}'}{2(V/2 + \tilde{V})} \delta \tilde{g}_{ij}$ has the same form as $g_{ij}$ in eq. (\ref{kinMetric}),
with $t$ and $\tilde{t}$ replaced by $V$ and $\tilde{V}$:
\begin{align}
\delta \tilde{g}_{ij} =
\left(
\begin{matrix}
(1 + \frac{V'}{\tilde{V}'})  u_{xx} + u_{yy} &  2u_{xy} \\
2 u_{xy} & (1 + \frac{V'}{\tilde{V}'}) u_{yy} + u_{xx} \\
\end{matrix} \right).
\end{align}
$V' = a\frac{\partial V}{\partial r}|_{r = a}$ and $\tilde{V}' = \sqrt{2} a \frac{\partial \tilde{V}}{\partial r} |_{r = \sqrt{2}a}$,
where $V(r)$ and $\tilde{V}(r)$ are the nearest and next-nearest neighbor interactions, which only depend
on the distance $r$ between the nearest or next-nearest neighbor sites. As will be discussed below,
the phonon Hall viscosity is proportional to the gravitational Hall viscosity only
when the two metrics, $g_{ij}$ and $\tilde{g}_{ij}$ are the same. %When these are directly proportional to each other,
%%we have
%the two metrics are the same - $\tilde{g}_{ij} = g_{ij}$, where $g_{ij}$ is given in (\ref{kinMetric}) -
%%the gravitational response is directly related to the phonon response. In particular,
%%for isotropic systems
%the gravitational Hall viscosity is proportional to the phonon Hall viscosity.
%Surprisingly, as we discuss below, we find also other limits in which phonon Hall viscosity is directly proportional to gravitational Hall viscosity.

To study the simplest possible scenario, we assume the electrons hop among a single s-wave orbital of the atoms.
The BCS mean-field Hamiltonian is then
\begin{equation}
H_{BCS} = \sum_{\bf k} (\epsilon_{\bf k}-\mu) c^\dagger_{\bf k} c_{\bf k} + \frac12 \sum_{\bf k} [\Delta_{\bf k} c_{{\bf k}}^\dagger c_{-\bf k}^\dagger + H.c.],
\end{equation}
where $\epsilon_k$ is given by Eq.\eqref{kinEn}.
For $p_x + ip_y$ pairing, we take the order parameter to be,
\begin{align}
\Delta_{\bf k} = &\Delta (\sin k_x + i\sin k_y)
\nonumber \\
& + \tilde{\Delta} (\sin k_x \cos k_y + i \sin k_y \cos k_x).
\end{align}
The order parameter must satisfy a self-consistency equation:
\begin{equation}
\Delta_{\bf k} = -\sum_{{\bf k}'} V_{{\bf k}{\bf k}'} \frac{\Delta_{{\bf k}'}}{2E_{{\bf k}'}},
\label{EQ:selfConsist}
\end{equation}
where $V_{{\bf k}{\bf k}'}$ is the Cooper channel of the interaction and $E_{\bf k} = \sqrt{(\epsilon_{\bf k}-\mu)^2 + |\Delta_{\bf k}|^2}$.
%Here, we encounter a subtlety in obtaining the phonon response: the self-consistency equation needs to be solved for every value of the strain parameters $u_{ij}$. However, in general there is no unique way to do this, even while imposing continuity on the solutions, which signals a shortcoming of the mean-field description.
%SB comment: I do not agree with this comment - I believe self-consistency condition does give us the response to the strain, and it fails only with respect to rotation.

\begin{figure}[tb]
\centerline{
\includegraphics[width=3.4in,height=2.3in]{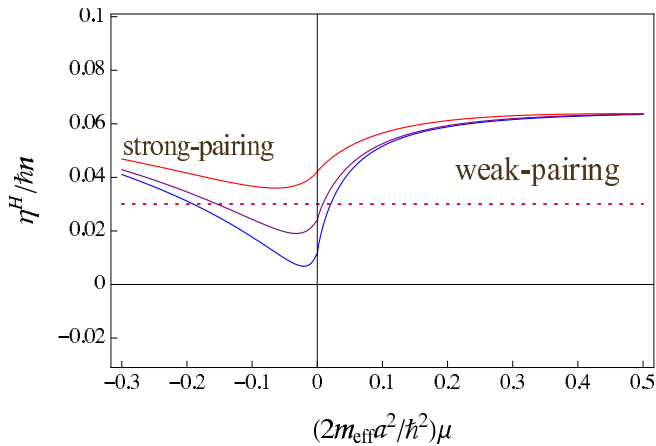}
}
\caption{The $p_x+ip_y$ superconductor phonon Hall viscosity. The filled curves are for the cases with $g_{ij} \neq \tilde{g}_{ij}$, while the dotted curves are for %$\eta^H_0$ is the Hall viscosity for the case
$g_{ij} = \tilde{g}_{ij}$. We have set $V'/V = 2t'/t$, $\tilde{t}/t=-\tilde{t}'/t = 0.75$, and $\tilde{V}/V=-\tilde{V}'/V = 0.40$. The pairing gap was set to have a Gaussian cutoff $\Delta = \hat{\Delta}k \exp[-k^2/2(\delta k)^2]$, with $\delta k = 1/a$ and $\hat{\Delta}(2m_{eff}a/\hbar^2) = 0.3, 0.4, 0.6$ for the blue, purple and red curves respectively. Note that while $\eta^H$ converges to the same asymptotic values in the weak-pairing limit, in the strong-pairing phase and near the quantum phase transition, it is not proportional to the density $n$ when $g_{ij} \neq \tilde{g}_{ij}$.}
\label{p+ip}
\end{figure}

To calculate the phonon Hall viscosity for this system, we need to obtain the effect of the lattice deformation
on the order parameter $\Delta_k$. For simplicity, we consider the long-wavelength continuum limit, where
\begin{align}
H_{BCS} = \frac12 \sum_k \Psi_k^\dagger H_{BdG}(k) \Psi_k,
\end{align}
where $\Psi_k^\dagger  = (c_k^\dagger, c_{-k})$, and
\begin{equation}
\hat{H}_{BdG} = \left[\begin{array}{cc} \epsilon_k-\mu & \Delta(|k|) (\hat{k}_i e_{xi} + i \hat{k}_j e_{yj})\\
                \Delta^*(|k|) (\hat{k}_i e_{xi} - i \hat{k}_j e_{yj}) & -\epsilon_k +\mu \end{array}\right].
%\equiv {\bf d}\cdot{\bm \tau},
\label{EQ:deformBdG}
\end{equation}
$\hat{k}_i$ is a unit vector, and $\Delta(|k|)$ is chosen to fall to zero far away from the Fermi surface (the cutoff scheme is explain in Fig.\ref{p+ip}), %. In our calculations,
%we will choose
%\begin{align}
%\Delta(k) = \hat{\Delta} k e^{- (k - k_F)^2/2k_F s^2},
%\end{align}
%where $\hat{\Delta}$ is a complex number, $k_F = \sqrt{2 m_{eff} \mu}$, $m_{eff}$ is the effective mass of the electrons,
$\mu$ is their chemical potential, and
%$s$ is a parameter that tunes the scale of the Gaussian cutoff around the Fermi surface.
$e_{ij} = \delta_{ij} + \delta e_{ij}$ where $\delta e_{ij}$ is a linear combination of the lattice distortions $w_{kl} \equiv \partial_k u_l$.
It is convenient to define an ``order parameter metric'' $g^{\Delta}_{ij} = e_{ia} e_{ja}$, which we can fix in terms of $g$ and $\tilde{g}$ using the
BCS self-consistency equation. In the continuum limit, where the system is rotationally invariant, it is simple to show (see Appendix \ref{BCSselfConsist}) that
\begin{equation}
g^\Delta_{ij} = \gamma g_{ij}  + (1-\gamma) \tilde{g}_{ij},
\end{equation}
where $\gamma$ is a constant that can be determined from the self-consistency equations (see Appendix \ref{BCSselfConsist}). To first order
in $w_{ij}$, we have $g_{ij}^\Delta = \delta_{ij} + (\delta e_{ij} + \delta e_{ji})$, so the above equation does not fix $\delta e_{ij} - \delta e_{ji}$.
We may fix $\delta e_{ij} - \delta e_{ji}$ by observing that the only affect of a rigid rotation of the crystal should be to rotate $k_x$ and $k_y$
into each other. Thus: $e_{ij} - e_{ji} = 2 m_{ij}$. These considerations fix the dependence of the order parameter on the lattice deformations.
Thus we can now use the Kubo formula and explicitly obtain the Hall viscosity:
\begin{align}
\eta_{ijkl} = \frac{1}{2} \frac{\hbar}{8 \pi^2} \int d^2k \hat{d} \cdot (\frac{\partial \hat{d}}{\partial w_{ij}} \times \frac{\partial \hat{d}}{\partial w_{kl}}) + (i \leftrightarrow k)
\end{align}
In Appendix \ref{BCSvisc}, we present some details of the calculation.
In the case where $\tilde{g}_{ij} = g_{ij}$, the calculation simplifies considerably and we find a simple result
\begin{align}
\eta^H = \frac{\tilde{t}'t'}{4(t/2+\tilde{t})^2} \frac14\hbar n=\frac{\tilde{t}'t'}{4(t/2+\tilde{t})^2}\eta^H_{gr},
\end{align}
where the gravitational Hall viscosity is $\eta^H_{gr} = \frac{1}{4}\hbar n$. The constant of proportionality
$\eta^H/\eta^H_{gr}$ is typically of order one. In Fig. \ref{p+ip}, we show the Hall viscosity calculated from the results presented in Appendix \ref{BCSvisc} for $g_{ij} \neq \tilde{g}_{ij}$. While this does not modify the Hall viscosity in the weak pairing limit much, the behavior close to the weak to strong pairing transition is dependent on completely non-universal features, such as the frequency dependence of the pairing gap.

For a system with a circular Fermi surface, the $p_x + i p_y$ state has another interesting feature:
the $U(1)_\phi \times U(1)_{L_z}$ symmetry associated with particle number and angular momentum conservation
is spontaneously broken to a diagonal subgroup, $U(1)_{L_z - \phi}$. This implies that in the effective action,
$m_{xy}$ and $\phi$, %where $\phi$ is
the angle through which the crystal is rotated about the $z$ direction and the overall phase of the order parameter, respectively, should appear together as $m_{xy} + \phi$. %since $m_{xy}$ is the angle through which the crystal is rotated about the $z$ direction.
This means that the effective action of the crystal involves the phase of the order parameter as well:
\begin{align}
\delta S_H =  2\int d^d x dt &[\eta^H (u_{xx} - u_{yy}) \dot{u}_{xy}\nonumber\\
&+ \eta^M (u_{xx} + u_{yy}) (\dot{m}_{xy} + \dot{\phi})].
\end{align}
Physically the $U(1)_{L_z - \phi}$ symmetry requires $L_z$ to increase by $\hbar$ when two electrons are adiabatically added. Since $m_{xy}$ and $\phi$ are conjugate variables to $L_z$ and the number of Cooper pairs, the above effective action contains the Berry phase term for this adiabatic process:
\begin{equation}
\hbar \frac{\delta n}{2} = \frac{\partial \mathcal{L}_H}{ \partial \dot{\phi}} = \frac{\partial \mathcal{L}_H}{ \partial \dot{m_{xy}}} = \delta L_z.
\end{equation}
The physical consequences of the coupling between the uniform compression, $u_{xx} + u_{yy}$, and
$\phi$ means %is as follows. Since $\theta$ and particle number, $N$, are conjugate, it follows
that the change in particle density is given by
\begin{align}
\delta n =  4 \eta^M (u_{xx} + u_{yy}) = 4 \eta^M \frac{\delta A}{A},
\end{align}
where $A$ is the area of the 2D system and $n$ is the particle density, when $\mu$ is held constant \footnote{Read \cite{Read09, ReadRezayi10} considered the case with constant particle number rather than constant chemical potential; hence the absence of $\dot{m}_{xy} + \dot{\phi}$.}. A generic superconductor of course has
\begin{align}
\delta n / n = \alpha(\mu) \delta A/A,
\end{align}
where $\alpha(\mu)$ is a constant depending on the chemical potential; %. ({\bf explain more about what is $\alpha(\mu)$})
note that we would have $\alpha = -1$ if the total particle number had been fixed. The additional symmetry in this problem, which relates
phase rotations to rotations of the crystal, sets $\alpha(\mu) = -4 \eta^M/n$. This suggests a way to experimentally
measure $\eta^M$ in systems that have an additional symmetry involving spatial rotations of the crystal.

\section{Physical conseqeunces of phonon Hall viscosity}\label{sec:observation}

\subsection{Acoustic phonon dynamics}

Consider the effective long-wavelength elasticity theory of a crystal, given by (\ref{phAction}) and (\ref{phHallAction}). 
Note that this is an expansion in the displacement fields and its gradients. Since for the sound
waves $\omega \propto |k| + \cdots$, the Hall viscosity terms are actually of order $k^3$, so for consistency
one must also include a term of the form $\delta S_3 = \int d^dx dt \lambda_{ijklm} \partial_m \partial_i u_j \partial_k u_l$, but
such a term vanishes in the presence of inversion symmetry. As we noted in Section \ref{effActionSec}, anharmonic effects
are $\mathcal{O}(k^4)$, so Hall viscosity may be distinctly measurable because its effects appear at lower order in $k$. We briefly mention
the effects of impurities later. The physical consequences of Hall viscosity terms can be analyzed most simply by considering 2D systems 
whose long-wavelength elastic theory is isotropic. This would be directly physically relevant for 2D systems with
square lattice symmetry; the considerations directly apply for layered 3D crystals as well, where the 2D layers have
a square lattice symmetry and where we consider phonons with wave-vector oriented parallel to the 2D layers. For such systems, 
the elastic theory simplifies and one obtains for the equation of motion:
\begin{align}
\label{eom}
\ddot{u}_i = c_t^2 \nabla^2 u_i + (c_l^2  -c_t^2) \partial_i \nabla \cdot \v{u} + \eta \nabla^2 \epsilon^{ij} \dot{u}_j/\rho,
\end{align}
where the indices $i$, $j$ run over the 2D spatial coordinates,  $c_t$ and $c_l$ are the transverse and longitudinal sound
velocities, respectively, and $\eta \equiv \eta_{xxxy}$ is the Hall viscosity.
It is simple to show that such a wave equation does not admit purely transverse or purely longitudinal
solutions. Let us denote $e_{\pm}$ as the eigenmodes of the system, and let the basis
$\left( \begin{matrix} 1 & 0 \end{matrix} \right)^T$ and 
$\left(\begin{matrix} 0 & 1 \end{matrix} \right)^T$ correspond to the longitudinal and transverse
acoustic phonon modes, respectively. In this basis, the eigenmodes of the system in the presence of the Hall viscosity are
\begin{align}
e_+ \propto \left( \begin{matrix}
1 \\ -i x
\end{matrix} \right) + \mathcal{O}(x^2), \;\;\;\;\;
e_- \propto
\left( \begin{matrix}
-ix \\ 1 \end{matrix} \right) + \mathcal{O}(x^2),
\end{align}
where $x \equiv \frac{\omega}{\omega_v} = \frac{\eta \omega}{\rho (c_t^2 - c_l^2)}$ is a dimensionless parameter.
This defines the characteristic frequency $\omega_v = \frac{\rho (c_t^2 - c_l^2)}{\eta} \sim \frac{B}{\eta}$, where
$B$ is the bulk modulus of the crystal. (Note that for a crystal, $c_l > \alpha c_t$, for some constant $\alpha$ of order unity, 
which is why $\rho (c_t^2 - c_l^2) \sim \rho c_l^2 \sim B$). Observe that to
linear order, there is a $\frac{\pi}{2} \sgn(\eta)$ phase shift between the longitudinal and transverse modes.
The dispersion relation is
\begin{align}
\omega^2 = &\frac{k^2}{2} [c_l^2 + c_t^2 + \eta^2 k^2/\rho^2 \pm
\nonumber \\
&\sqrt{c_l^4 + (c_t^2 + k^2 \eta^2/\rho^2)^2 + 2c_l^2 (k^2 \eta^2/\rho^2 - c_t^2)  }].
\end{align}
The shift in frequency for a given acoustic phonon mode for finite $\eta$ is 
$\Delta \omega/ \omega(\eta = 0) \sim x(\omega (\eta = 0))^2$.
In principle then the shift in frequency can determine $\eta$. However, since this is not sensitive
to the sign of the Hall viscosity, it may not be a useful method in practice for determining the Hall
viscosity. 
%With increasing $\eta$, one finds that the longitudinal mode at $\eta = 0$, with
%$\omega(\eta = 0) = c_l |k|$, evolves into a mode at finite $\eta$ whose dispersion becomes
%quadratic when $k \eta/\rho c \gg 1$; in the quadratic limit, these are the ``viscosity waves'' considered in
%\Ref{Avron98}; however in this limit the effective theory above should not be taken seriously; one would need to include
%the higher derivative terms $\delta S_{4} = \lambda_{ijklm} \partial_m u_{ij} \partial_n u_{kl}$ to be consistent at this order.
%The transverse mode at $\eta = 0$ evolves into a mode whose dispersion asymptotes to a constant for
%large $k \eta/\rho c$.

An analysis of surface (Rayleigh) waves of a 3D medium with non-zero $\eta_{xxxy} = \eta_{yyyx}$ displays similar behavior.
For a medium with surface at $z = 0$, a surface wave travelling in the $x$-direction must have
$u_y = 0$ in the absence of Hall viscosity, due to stress-free boundary conditions. In the presence
of a Hall viscosity, the surface wave acquires a $u_y$ component, which to linear order in
$\eta \omega/\rho c^2$ differs by a phase shift of $\pi/2$ and has a relative amplitude of $\eta \omega/\rho c^2$.

\subsection{Numerical estimates and discussion of possible experimental detection }
\label{expDisc}

As explained above, the physical consequences of a Hall viscosity in the phonon effective action is the mixing of longitudinal and
transverse sound modes, which at a frequency $\omega$ is determined by $\omega/\omega_v$, with
$\omega_v = \rho c^2/\eta^H$ a characteristic frequency scale associated with the Hall viscosity.
To lowest order in $\omega/\omega_v$, there is a phase shift of $\pi/2$. For the bulk modes,
a more precise value of $\omega_v$ is $\omega_v = \rho (c_l^2 -c_t^2)/\eta^H$, and the amplitude
ratio between the two modes in the elliptical polarization is $\omega / \omega_v$. It is not clear what the
minimal experimentally measurable value of the Hall viscosity is. From elementary considerations, we can put a
rough bound on what may realistically be measured. First, for the sound waves to not
destroy the crystal, we expect that the strain is small: $|\partial u| \ll 1$, which implies that $|k u| \ll 1$.
Since we have roughly $\omega \sim c k$, this implies $ \omega u \ll c$. The amount of mixing is determined by
$\omega/\omega_v$, so the amount of amplitude from the other mode that is mixed in is $(\omega/\omega_v) u$.
For this to be realistically measurable, this should be much larger than the size of the quantum fluctuations of the
wave function of an atom, which is on the order of 0.1 \AA. That is, $|\omega u| \gg \omega_v \times 0.1 \AA$.
Thus we have the conditions $ \omega_v \times 0.1 \AA \ll u \ll c$.
For a typical sound velocity of $5 \times 10^5$  cm/s, this implies
\begin{align}
\omega_v \ll 5 \times 10^{14} s^{-1}.
\end{align}
For smaller sound velocities, this bound will be smaller. This is not a fundamental bound, but a practical one. This is
because in principle it is possible to measure oscillations of the center of the wave function of an atom at a
resolution that is smaller than the characteristic size of its wave function.

The two-dimensional mass density of the crystal is $\rho \propto A m_p/a^2$, where $A$ is the atomic number of atoms of the crystal.
Typically, $A \sim 10$, and $a \sim 4 \times 10^{-8} cm$, so $\rho \sim 10^{-8} g/cm^2$.
Furthermore, $c_l^2 - c_t^2 \sim (\alpha^2 - 1) 10^{10} cm^2/s^2$,
where $\alpha = c_l/c_t$ and typically $\alpha \sim 2$. $\eta^H \sim
\eta^H_{gr} \sim \hbar n_e$, where $n_e$ is the electron density. Thus, for a typical 2D electron density
of $10^{15} cm^{-2}$, with $\alpha \sim 2$, we see $\omega_v \sim 10^{14} s^{-1}$. For a 1 GHz measurement,
$\omega/\omega_v \sim 10^{-5}$; at 100 GHz, $\omega/\omega_v \sim 10^{-3}$ and, depending on material parameters,
could be closer to $10^{-2}$. Note that the frequencies must be much less than the energy gap of the electronic state, which for a
10 K gap translates to approximately 2 THz, and also much less than the phonon Debye frequency, which is close to 10 THz.

For quantum Hall states induced by an external magnetic field, the electron densities are usually low,
$n_e \sim 10^{11} cm^{-2}$, yielding immeasurably small values for $\omega/\omega_v$.
An exception may be graphene, where recent advancements in applying extremely large gate
voltages may allow for much larger densities of electrons participating in quantum Hall states.\cite{EK1005}
The necessary values of $n_e \sim 10^{15} cm^{-2}$ usually appear in states that spontaneously
break time-reversal symmetry, such as quantum anomalous Hall states, ferromagnetic insulators,
or chiral superconductors, where the effective magnetic moment per lattice site is much larger than
could be produced by an external magnetic field.

The effects discussed here would most easily be measurable in bulk, layered 3D crystals, for phonons
propagating along an in-plane high-symmetry direction. While there are
a number of examples of 3D IQH states, the value of the Hall viscosity is probably too small to be measured,
since the particle density is too small, though not typically as low as in 2D quantum wells. More promising
systems are those that spontaneously break time-reversal symmetry, because those typically will have
much higher angular momentum densities.  One promising candidate may be the chiral superconductor
Strontium Ruthenate, which exists as a 3D crystal and may have a large enough Hall viscosity because it
spontaneously breaks time-reversal. Another promising set of materials to measure a phonon Hall viscosity
are 3D ferromagnetic insulators, for example those discussed in \Ref{ferroSpinels65,GdNferroInsRef, Hulliger} .
Note that in cases where the spin gap is small, an external magnetic field can be used to ensure the electronic state
is fully gapped.

In principle, one way to measure such an effect would be through pulsed echo ultrasound meaurements, which have
been successful in detecting circular polarization between transverse sound waves. However, while bulk pulsed ultrasound
seems to be limited to frequencies on the order of 1 GHz, it is not clear what the ultimate bounds are on an
experimentally accessible amplitude ratio between transverse and
longitudinal waves. A more promising experimental technique appears to be time-dependent
x-ray diffraction \cite{TrigoReis2010}. Such techniques have been developed only recently over the last decade and have been
used to directly image acoustic phonon modes \cite{RD0172,LK0011}.

One complication of measuring the phonon Hall viscosity is related to the effects of crystal disorder, which can also
mix transverse and longitudinal waves. However, the effects of disorder are not sensitive to the sign of the time-reversal symmetry
breaking of the electronic state; this dependence on the sign of the time-reversal symmetry breaking is unique to the Hall viscosity, and can
be used to extract the phonon Hall viscosity even for imperfect crystals.

We would like to point out that related phenomena occur in various other time-reversal breaking systems. The
phonons in a ferromagnet, for example, can exhibit acoustic Faraday rotation, where the two transverse modes acquire
a circular or elliptic polarization\cite{Kittel58,ML6297}. However such systems cannot be described by a simple local effective action
in terms of the strain fields because they are coupled to magnons, which are gapless; integrating out the magnons will
result in non-local terms in the crystal effective action. The physical manifestations of such phenomena are also quite different;
they occur as resonances when the frequency and wavelength of the phonons and magnons are matched. More directly related
phenomena have been considered in the case of ionic crystals in an external magnetic field \cite{Vineyard85}, and in Tkachenko
modes of vortex lattices in rotating superfluids \cite{Tkachenko1, Tkachenko2, BaymChandler83}. In these situations, one obtains
a related equation of motion as in (\ref{eom}).

\section{Conclusion}

We have proposed the acoustic phonon Hall viscosity as a novel probe into the adiabatic Berry curvature of the
many-body electron wave function for gapped, time-reversal symmetry breaking electronic states, and
we have computed it and studied its behavior in a number of theoretical models. In some simple cases, we have found
that the phonon Hall viscosity is proportional, with a numerical factor of order 1, to the gravitational Hall viscosity
of the continuum electronic theory. Our numerical estimates indicate that this is a measurable effect
and there may be a number of materials, particularly the ferromagnetic insulators, which might be suitable
candidates for experimentally detecting the phonon Hall viscosity by measuring time-reversal symmetry breaking corrections
to acoustic phonon dynamics. Since the effects are expected to be small and their measurement would require high spatial resolution,
it appears that time-dependent x-ray diffraction may be currently the most promising probe. As phonon Hall viscosity is developed into
a more mature experimental probe, we hope that it can eventually be useful as a novel lens into the possible
topological behavior of electron systems.

%The most promising situations in which the Hall viscosity may be experimentally measurable occur in systems that
%spontaneously break time-reversal symmetry. However such systems also often come with gapless spin waves, so the electron system
%is not completely gapped. An important future direction would be to develop a theory of phonon Hall viscosity coexisting with
%gapless spin waves; in such situations, the calculation of the phonon Hall viscosity presented in this paper must be modified, at least
%for frequencies close to the magnon-phonon resonance.

We thank A. Auerbach, T. Deveraux, T. Hughes, S. Kivelson, R.B. Laughlin, S. Riggs, D. T. Son, J. Tranquada and C. Varma for helpful discussions. 
This work is supported by the Alfred P. Sloan Foundation (XLQ), the Simons Foundation (MB), and the DOE under contract DE-AC02-76SF00515 (SBC) .
We thank the Aspen Center for Physics and the KITP for hospitality while this work was being completed. 

\appendix

\section{$p+ip$ BCS self-consistency}
\label{BCSselfConsist}

The BCS mean-field Hamiltonian is:
\begin{equation}
H_{BdG} = \sum_{\bf k} (\epsilon_{\bf k}-\mu)c^\dagger_{\bf k} c_{\bf k} - \sum_{\bf k} \Delta_{\bf k} c_{-{\bf k}} c_{\bf k},
\end{equation}
where
\begin{align}
\Delta_{\bf k} = &\Delta (\sin k_x + i\sin k_y)
\nonumber \\
&+ \tilde{\Delta} (\sin k_x \cos k_y + i \sin k_y \cos k_x)
\end{align}
and the kinetic energy is that of Eq.\eqref{kinEn}. $\Delta_{\bf{k}}$ satisfies a self-consistency equation:
\begin{equation}
\Delta_{\bf k} = -\sum_{{\bf k}'} V_{{\bf k}{\bf k}'} \frac{\Delta_{{\bf k}'}}{2E_{{\bf k}'}},
\label{EQ:selfConsistApp}
\end{equation}
where $V_{{\bf k}{\bf k}'}$ is the Cooper channel of the interaction and $E_{\bf k} = \sqrt{(\epsilon_{\bf k}-\mu)^2 + |\Delta_{\bf k}|^2}$.

When the crystal is strained, the order parameter will take the following form in the continuum limit:
\begin{align}
\Delta_k = \Delta (\hat{k}_i e_{xi} + i \hat{k}_i e_{yi}),
\end{align}
where $e_{ab} [\{\partial_i u_j\}]$ are functions of the distortion tensor $w_{ij} \equiv \partial_i u_j$.
In order to obtain the phonon response, we need to obtain this function to linear order in $\partial_i u_j$. First, observe that due to the U(1)$_{L_z} \times$ U(1)$_\phi \to$ U(1)$_{L_z - \phi}$ symmetry breaking, $\phi$ spatial rotation is equivalent to the gauge transformation $\Delta \to \Delta \exp(i\phi)$. Therefore %under only a 90 degree rotation of the crystal we expect $k_x \rightarrow k_y$ and $k_y \rightarrow - k_x$, so that $\Delta_k \rightarrow \Delta (k_y - i k_x)$. Thus it follows that
\begin{align}
e_{xy} - e_{yx} = w_{xy} - w_{yx}.
\end{align}
Next, observe that $e_{xx}$, $e_{yy}$, and $e_{xy} + e_{yx}$ are symmetric in
$x$ and $y$, so they can only depend on the strain tensor $u_{ij}$. %Define $\tilde{g}_{ij} \equiv \frac12 (\tilde{e}_{ij} + \tilde{e}_{ji})$.
The self-consistency equation can be thought of as a constraint on the ``order parameter metric" $g^\Delta$:
\begin{align}
f (g^\Delta, g, \tilde{g}) = 0.
\end{align}
Considering the variations of this:
\begin{align}
\delta g^\Delta_{ij} \frac{\partial f}{\partial g^\Delta_{ij}} + \delta g_{ij} \frac{\partial f}{\partial g_{ij}} + \delta \tilde{g}_{ij} \frac{\partial f}{\partial \tilde{g}_{ij}}  = 0.
\end{align}
For a rotationally invariant system, $\partial f/\partial g_{ij} \propto \delta_{ij}$, and similarly for
$\partial f/\partial \tilde{g}_{ij}$. This implies that $\delta g^\Delta_{ii} = \gamma \delta g_{ii} + \tilde{\gamma} \delta \tilde{g}_{ii} $,
where $\gamma$ and $\tilde{\gamma}$ are constants. Now observe that when $g_{ij} = \tilde{g}_{ij}$,
we are merely implementing a coordinate transformation, so we should have $g^\Delta_{ij} = g_{ij}$,
which implies $\gamma + \tilde{\gamma} = 1$. Furthermore, for a rotationally invariant system,
the deformations $\delta g_{xx} = - \delta g_{yy} = e$ and $\delta \tilde{g}_{xx} = - \delta \tilde{g}_{yy} = \tilde{e}$ are
equivalent to the deformations $\delta g_{xy} = e$ and $\delta \tilde{g}_{xy} = \tilde{e}$, because for a rotationally invariant system,
the two types of deformations simply differ by a rotation.
%Observe that $\partial f/\partial g_{ij}$ is a symmetric $2\times 2$ matrix. For a rotationally invariant system,
%$\partial f/\partial g_{ij} \propto \delta_{ij}$. Lastly, we note that
%$\partial f/\partial g^\Delta_{ij} + \partial f/\partial g_{ij} + \partial f/\partial \tilde{g}_{ij} =0$,
%for when we have $g_{ij} = \tilde{g}_{ij}$, we are merely implementing coordinate
%transformation for the order parameter and therefore %we should have
Thus, we conclude:
\begin{align}
g^\Delta_{ij} = \gamma g_{ij} + (1-\gamma) \tilde{g}_{ij}.
\end{align}
The constant $\gamma$ can be found from the self-consistency equation.

To actually calculate $\gamma$ from the self-consistency equation, we note that the
assumption we made above tells us that the effect of change in the kinetic metric $g_{ij}$
should be proportional to the effect of coordinate transformation $\delta_{ij} \to g_{ij}$.
This means that, if we consider the case $\delta g_{xx} = -\delta g_{yy} = e_1$ and $\delta \tilde{g}=0$,
the change in the order parameter should come out as $\delta \Delta_{\bf k} = \gamma e_1 \Delta (\hat{k}_x -  i \hat{k}_y)/2$.
Thus, to the self-consistency condition
\begin{widetext}
\begin{align}
\frac{\partial \Delta_{\bf k}}{\partial e_1} = -\sum_{{\bf k}'}V_{{\bf k}{\bf k}'}\left[\frac{1}{2E_{{\bf k}'}}\frac{\partial \Delta_{{\bf k}'}}{\partial e_1}%\right.\nonumber\\
%&\left.
-\frac{\Delta_{{\bf k}'}}{2E^2_{{\bf k}'}}\left(\frac{\epsilon_{{\bf k}'}-\mu}{E_{{\bf k}'}}\frac{\partial \epsilon_{{\bf k}'}}{\partial e_1} + \frac{|\Delta_{{\bf k}'}|}{E_{{\bf k}'}}\frac{\partial |\Delta_{{\bf k}'}|}{\partial e_1}\right)\right],
\label{EQ:selfConsistApp2}
\end{align}
we can insert
\begin{align}
\frac{\partial \Delta_{\bf k}}{\partial e_1} =& \frac{1}{2}\gamma \Delta (\hat{k}_x -  i \hat{k}_y),\nonumber\\
\frac{\partial \epsilon_{\bf k}}{\partial e_1} =& \epsilon_{\bf k}(\hat{k}_x^2 - \hat{k}_y^2),
\label{EQ:metricTransf}
\end{align}
to obtain
\begin{align}
\frac{1}{2}\gamma \Delta (\hat{k}_x -  i \hat{k}_y) = -\frac{1}{2}\gamma \sum_{{\bf k}'}V_{{\bf k}{\bf k}'}\frac{\Delta (\hat{k}'_x -  i \hat{k}'_y)}{2E_{{\bf k}'}}+\sum_{{\bf k}'}V_{{\bf k}{\bf k}'}\frac{\Delta_{{\bf k}'}\epsilon_{{\bf k}'}(\epsilon_{{\bf k}'}-\mu)}{2E^3_{{\bf k}'}}(\hat{k}_x^2 - \hat{k}_y^2)+\sum_{{\bf k}'}V_{{\bf k}{\bf k}'}\frac{\Delta_{\bf k'} |\Delta_{\bf k}'|}{2E^3_{{\bf k}'}}\frac{\partial |\Delta_{{\bf k}'}|}{\partial e_1}
\label{EQ:selfConsistApp3}
\end{align}
and solve for $\gamma$. For this step, it is convenient to eliminate $\partial |\Delta_{{\bf k}'}|/\partial e_1$ through a coordinate transformation argument, which
makes use of the fact that in the integral over $k'$, we can do a coordinate transformation
$(k'_x, k'_y) \to (k'_x, k'_y) + \gamma e_1 (k'_x, -k'_y)/2$ without changing the value of the integral because the Jacobian of this transformation is
one, to linear order in $e_1$.
%makes use of the fact that in Eq.\eqref{EQ:selfConsistApp}, $\Delta_{\bf k}$ will remain invariant when we apply the $(k'_x, k'_y) \to (k'_x, k'_y) + e_1 (k'_x, -k'_y)/2$ %transformation on $V_{{\bf k}{\bf k}'}, \epsilon_{{\bf k}'}$ and $\Delta_{{\bf k}'}$.
This gives us
\begin{align}
\sum_{{\bf k}'}V_{{\bf k}{\bf k}'}\frac{\Delta_{\bf k'} |\Delta_{\bf k}'|}{2E^3_{{\bf k}'}}\frac{\partial |\Delta_{{\bf k}'}|}{\partial e_1}= \sum_{{\bf k}'}V_{{\bf k}{\bf k}'}\left(\frac{1}{2E_{{\bf k}'}}\frac{\partial \Delta_{{\bf k}'}}{\partial e_1} -\frac{\Delta_{{\bf k}'}}{2E^2_{{\bf k}'}}\frac{\epsilon_{{\bf k}'}-\mu}{E_{{\bf k}'}}\frac{\partial \epsilon_{{\bf k}'}}{\partial e_1}\right)-\sum_{{\bf k}'}\frac{\partial V_{{\bf k}{\bf k}'}}{\partial e_1}\frac{\Delta_{{\bf k}'}}{2E_{{\bf k}'}}.
\label{EQ:coordTransf}
\end{align}
We can %solve for
obtain the derivatives on the right-hand side by noting that this coordinate transformation gives us
\begin{equation}
\Delta_{{\bf k}'} \to \Delta_{{\bf k}'} + \frac{1}{2}\gamma \Delta (\hat{k}'_x -  i \hat{k}'_y),\,\,\,\,\,\epsilon_{{\bf k}'} \to \epsilon_{{\bf k}'}+\gamma\epsilon_{\bf k}(\hat{k}_x^2 - \hat{k}_y^2),\,\,\,\,\,V_{{\bf k}{\bf k}'} \to V_{{\bf k}{\bf k}'}-\gamma V(\hat{k}_x\hat{k}'_x - \hat{k}_y\hat{k}'_y)
\end{equation}
where we used $V_{{\bf k}{\bf k}'} = -2V(\hat{\bf k}\cdot\hat{\bf k}')$. Since the change in $|\Delta_{{\bf k}'}|$ is same, we can insert this result from the coordinate transformation into Eq.\eqref{EQ:selfConsistApp2}. %Combining Eqs.\eqref{EQ:selfConsistApp2} and \eqref{EQ:coordTransf}, we obtain
%In order to insert the above equation into Eq.\eqref{EQ:selfConsistApp2}, we choose $\tilde{e}_1$ such that we have $\Delta_{{\bf k}'} \to \Delta_{{\bf k}'} + \gamma e_1 \Delta (\hat{k}_x -  i \hat{k}_y)/2$.
\end{widetext}

We find at the weak coupling limit
\begin{equation}
\gamma = -\frac{1- \ln (2\epsilon_c/|\Delta_0|)}{1+\ln (2\epsilon_c/|\Delta_0|)},
\end{equation}
where $\epsilon_c$ is the cutoff energy of the pairing and $|\Delta_0|$ is the value of $\Delta$ at the Fermi surface. The negative sign of $\gamma$ is because our kinetic metric deformation increases the density of state around $(0,\pm k_f)$ compared to $(\pm k_f, 0)$ and therefore it is energetically advantageous to have the $p_y$ pairing to be stronger than the $p_x$ pairing.
%\section{Hall viscosity and effect of rotation}

\section{p+ip BCS Hall viscosity calculation}
\label{BCSvisc}

%\begin{widetext}
Now we would like to calculate the Hall viscosity for the $p_x+ip_y$ BdG Hamiltonian.
We want: %$\eta^H = \eta_{xxxy} + \eta_{xxyx}$ where
\begin{align}
\eta^H = \frac{1}{2}(\eta_{xxxy} + \eta_{xxyx}) = \frac{\hbar}{16 \pi^2}  \int d^2k \hat{\bf d} \cdot \left(\frac{\partial \hat{\bf d}}{\partial w_{xx}} \times \frac{\partial \hat{\bf d}}{\partial u_{xy}}\right)
\end{align}
From
%\begin{align}
$H_{BdG} = \v{d} \cdot {\bm \tau}$, we have
%\nonumber \\
\begin{align}
d_x &= \Delta (\hat{k}_x e_{xx} + \hat{k}_y e_{xy}),
\nonumber \\
d_y &= \Delta (\hat{k}_y e_{yy} + \hat{k}_x e_{yx}),
\nonumber \\
d_z &= \frac{1}{2m^*}k_i k_j g_{ij} - \mu;
\end{align}
note that in a flat metric, $d_\parallel \equiv \sqrt{d_x^2 + d_y^2}$ and $d_z$ would depend only on $k$. In this section, since ${\bf d}_\parallel$ has explicit dependence only on $e_{ij}$ while $d_z$ has explicit dependence only on $g_{ij}$, we will treat $e_{ij}$ and $g_{ij}$ as independent variables. %In what follows, let us set $w_{ij} \equiv \partial_i u_j$.

For our calculation, we want to write the derivatives in terms of $g_{ij}$ and $e_{ij}$, which gives us
\begin{align}
\frac{\partial {\bf d}_\parallel}{\partial w_{xx}} =& \frac{1}{2}\left(\tilde{a}\frac{\partial {\bf d}_\parallel}{\partial e_{xx}} + \tilde{b}\frac{\partial {\bf d}_\parallel}{\partial e_{yy}}\right) = \frac{1}{2}\left(\tilde{a} k_x \frac{\partial {\bf d}_\parallel}{\partial k_x} + \tilde{b} k_y \frac{\partial {\bf d}_\parallel}{\partial k_y}\right),\nonumber\\
\frac{\partial d_z}{\partial w_{xx}} =& a\frac{\partial  d_z}{\partial g_{xx}} + b\frac{\partial d_z}{\partial g_{yy}} = \frac{1}{2}\left(a k_x\frac{\partial  d_z}{\partial k_x} + b k_y \frac{\partial d_z}{\partial k_y}\right),\nonumber\\
\frac{\partial {\bf d}_\parallel}{\partial u_{xy}} =& \tilde{b}\left(\frac{\partial {\bf d}_\parallel}{\partial e_{xy}}+\tilde{b}\frac{\partial {\bf d}_\parallel}{\partial e_{yx}}\right) = \tilde{b}\left(k_x \frac{\partial {\bf d}_\parallel}{\partial k_y}+ k_y \frac{\partial {\bf d}_\parallel}{\partial k_x}\right),\nonumber\\
\frac{\partial d_z}{\partial u_{xy}} =& 2b \frac{\partial d_z}{\partial g_{xy}} = b\left(k_x \frac{\partial d_z}{\partial k_y}+k_y \frac{\partial d_z}{\partial k_x}\right),
\end{align}
where
\begin{align}
a &= m_{eff}  \left(t' + \tilde{t}'\right),\nonumber\\
b &= m_{eff} \tilde{t}',\nonumber\\
\tilde{a} &= \gamma [m_{eff}(t' + \tilde{t}')] + (1-\gamma)\frac{V' + \tilde{V}'}{V/2 + \tilde{V}},\nonumber\\
\tilde{b} &= \gamma (m_{eff} \tilde{t}') + (1-\gamma)\frac{\tilde{V}'}{V/2 + \tilde{V}},
\end{align}
and we used
\begin{align}
\frac{\partial d_z}{\partial g_{ij}} =& \frac{1}{4}\left(k_i \frac{\partial d_z}{\partial k_j}+ k_j \frac{\partial d_z}{\partial k_i}\right)_{g_{ij}=\delta_{ij}},\nonumber\\
\frac{\partial {\bf d}_\parallel}{\partial e_{ij}} =& k_i \left.\frac{\partial {\bf d}_\parallel}{\partial k_j}\right\vert_{e_{ij} = \delta_{ij}}
\end{align}
To see this, first observe:
\begin{align}
\frac{\partial {\bf d}_\parallel}{\partial w_{ij}} &= \sum_{kl}\frac{\partial e_{kl}}{\partial w_{ij}}\frac{\partial {\bf d}_\parallel}{\partial e_{kl}},
\nonumber \\
\frac{\partial d_z}{\partial w_{ij}} &= \sum_{kl}\frac{\partial g_{kl}}{\partial w_{ij}}\frac{\partial d_z}{\partial g_{kl}}.
\end{align}
Assuming $w_{xy} - w_{yx} = e_{xy} - e_{yx}$, all $\partial g_{kl}/\partial w_{ij}, \partial e_{kl}/\partial w_{ij}$ vanish except for
%\begin{align}
%\eta_{xxxy} = \frac{\hbar}{8\pi^2} \frac{1}{a^2} f(\Delta, \mu, t),
%\end{align}
%where
%\begin{align}
%f = \int d^2k \frac{a k_x^4 + b k_x^2 k_y^2 + c k_y^4 - \mu (k_x^2 \alpha_{xxxx} \alpha_{yxxy} - k_y^2 \alpha_{yyxx} \alpha_{xyxy})}{[\Delta^2 k^2 + (k^2/2m - \mu)]^{3/2}}.
%\end{align}
\begin{align}
a =& \frac{\partial g_{xx}}{\partial w_{xx}} = \frac{\partial g_{yy}}{\partial w_{yy}},\nonumber\\
b=& \frac{\partial g_{xx}}{\partial w_{yy}} = \frac{\partial g_{yy}}{\partial w_{xx}} = \frac{\partial g_{xy}}{\partial w_{xy}} = \frac{\partial g_{xy}}{\partial w_{yx}},
\end{align}
and
\begin{align}
\tilde{a} =& 2\frac{\partial e_{xx}}{\partial w_{xx}} = 2\frac{\partial e_{yy}}{\partial w_{yy}},\nonumber\\
\tilde{b} =& 2\frac{\partial e_{xx}}{\partial w_{yy}} = 2\frac{\partial e_{yy}}{\partial w_{xx}}\nonumber\\
=& 2\frac{\partial e_{xy}}{\partial w_{xy}}-1 = 2\frac{\partial e_{xy}}{\partial w_{yx}}+1\nonumber\\
=& 2\frac{\partial e_{yx}}{\partial w_{xy}}+1 = 2\frac{\partial e_{yx}}{\partial w_{yx}}-1.
\end{align}
%with . %The constants $a$, $b$, $\tilde{a}$, and $\tilde{b}$ are:
%\end{widetext}
%We now have

Now we need to calculate
\begin{align}
\eta^H =& \frac{\hbar}{16\pi^2}\int d^2k \hat{\bf d} \cdot \left(\frac{\partial \hat{\bf d}}{\partial w_{xx}} \times \frac{\partial \hat{\bf d}}{\partial u_{xy}}\right)\nonumber\\
=& \frac{\hbar}{32\pi^2} ab \int d^2k \frac{1}{d^2}\hat{\bf d} \cdot \left[k_x\frac{\partial {\bf d}''}{\partial k_x}\times\left(k_x \frac{\partial {\bf d}'}{\partial k_y} + k_y \frac{\partial {\bf d}'}{\partial k_x}\right)\right]\nonumber\\
+&\frac{\hbar}{32\pi^2} b^2 \int d^2k \frac{1}{d^2}\hat{\bf d} \cdot \left[k_y\frac{\partial {\bf d}'}{\partial k_y}\times\left(k_x \frac{\partial {\bf d}'}{\partial k_y} + k_y \frac{\partial {\bf d}'}{\partial k_x}\right)\right]\nonumber\\
=& \frac{\hbar}{32\pi^2} ab \int k_x^2 d^2k \frac{1}{d^2}\hat{\bf d} \cdot \left(\frac{\partial {\bf d}''}{\partial k_x}\times \frac{\partial {\bf d}'}{\partial k_y}\right)\nonumber\\
+&\frac{\hbar}{32\pi^2} b^2 \int k_y^2 d^2k \frac{1}{d^2}\hat{\bf d} \cdot \left(\frac{\partial {\bf d}'}{\partial k_y}\times \frac{\partial {\bf d}'}{\partial k_x}\right)
\end{align}
where ${\bf d}' = (\tilde{b} {\bf d}_\parallel/b, d_z)$ and ${\bf d}'' = (\tilde{a} {\bf d}_\parallel/a, d_z)$. Then, using
\begin{align}
\frac{\partial}{\partial k_x} =& \cos \phi \frac{\partial}{\partial k} - \frac{\sin \phi}{k}\frac{\partial}{\partial \phi},\nonumber\\
\frac{\partial}{\partial k_y} =& \sin \phi \frac{\partial}{\partial k} + \frac{\cos \phi}{k}\frac{\partial}{\partial \phi},
\end{align}
%and
we obtain:
\begin{widetext}
\begin{align}
\eta^H %=& \frac{\hbar}{16 \pi^2}  \int d^2k \hat{\bf d} \cdot \left(\frac{\partial \hat{\bf d}}{\partial w_{xx}} \times \frac{\partial \hat{\bf d}}{\partial w_{xy}}\right)\nonumber\\
=& C_1 \int k^2 dk \frac{\partial \hat{d}_z}{\partial k} + C_2 \int k^2 dk \left[\hat{d}_z (1-\hat{d}_z^2)\frac{1}{d}\frac{\partial d}{\partial k} - \hat{d}^2_z \frac{\partial \hat{d}_z}{\partial k}\right]+ C_3 \int k^3 dk\left[(1-\hat{d}_z^2)\frac{1}{d}\frac{\partial d}{\partial k}-\hat{d}_z\frac{\partial \hat{d}_z}{\partial k}\right]\left(\frac{\partial \hat{d}_z}{\partial k}+\hat{d}_z\frac{1}{d}\frac{\partial d}{\partial k}\right)\nonumber\\
=& 8\pi C_1 n - \frac{2}{3} C_2 \int k dk (1-\hat{d}_z^3)+ C_2 \int k^2 dk \hat{d}_z (1-\hat{d}_z^2)\frac{1}{d}\frac{\partial d}{\partial k}+ C_3 \int k^3 dk\left[(1-\hat{d}_z^2)\frac{1}{d}\frac{\partial d}{\partial k}-\hat{d}_z\frac{\partial \hat{d}_z}{\partial k}\right]\left(\frac{\partial \hat{d}_z}{\partial k}+\hat{d}_z\frac{1}{d}\frac{\partial d}{\partial k}\right),
%\frac{1}{d^2}\left\vert\begin{array}{ccc}
%\hat{d}_\parallel \cos \phi & \hat{d}_\parallel \sin \phi & \hat{d}_z\\
%\frac{1}{2}k \frac{\partial d_\parallel}{\partial k} (\tilde{a} \cos^2 \phi + \tilde{b} \sin^2 \phi) \cos \phi + \frac{1}{2}d_\parallel (\tilde{a}-\tilde{b})\cos\phi\sin^2\phi & -\frac{1}{2}k \frac{\partial d_\parallel}{\partial k} (\tilde{a} \cos^2 \phi + \tilde{b} \sin^2 \phi) \sin \phi + \frac{1}{2}d_\parallel (\tilde{a}-\tilde{b})\cos^2\phi\sin\phi &
%\end{array}\right\vert
\label{EQ:integral}
\end{align}
%\end{widetext}
where
\begin{align}
C_1/\hbar =& \frac{1}{128\pi}(\tilde{a}b+3a\tilde{b}-4b\tilde{b}),\nonumber\\
C_2/\hbar =& C_1/\hbar -\frac{1}{32\pi}\tilde{b}(\tilde{a}-\tilde{b}),\nonumber\\
C_3/\hbar =& \frac{1}{256\pi}(a\tilde{b} - \tilde{a}b).
\end{align}
Now note that when the kinetic and interaction metrics are equal, we have $a = \tilde{a}$, $b = \tilde{b}$, which means $C_1/\hbar = b(a-b)/32\pi$ and $C_2 = C_3 =0$. Inserting these into Eq.\eqref{EQ:integral} gives us
\begin{equation}
\eta^H = 8\pi C_1 n = b(a-b) \hbar n/4 = (t/2 + \tilde{t})^{-2} \frac{t'\tilde{t}'}{4} \frac{\hbar n}{4}.
\end{equation}
%and it has been shown \cite{Read09, ReadRezayi10} that $\eta^H_{gr} = \hbar n/4$.
%\section{Effect of rotation on p+ip superconductor and QAH}
%When $\delta \tilde{g}_{ij} \propto \delta g_{ij}$, then we have $\tilde{b}/\tilde{a} = b/a$ and thus $C_3 = 0$. If this condition is satisfied in the weak pairing limit, where $\hat{d}_z^2 \to 1$ everywhere, $\eta^H \propto n$ is satisfied.

In the weak pairing limit, even when the kinetic and interaction metrics are different, we have $\eta^H \propto n$ under some reasonable assumptions. First, note that
\begin{align}
\int k dk (1-\hat{d}_z^3) =& \int k dk (1-\hat{d}_z) = 4\pi n,\nonumber\\
\int k^2 dk \frac{d_z d_\parallel^2}{d^5}\left(d_z\frac{\partial d_z}{\partial k}+d_\parallel\frac{\partial d_\parallel}{\partial k}\right) =& \int k dk \frac{d_z d_\parallel^2}{d^5} k \frac{\partial d_z}{\partial k} = k_f^2 \int_{-\mu}^\infty d\xi \frac{\xi^2 \Delta^2 }{(\xi^2 + \Delta^2)^{5/2}}= \frac{2}{3}k_f^2 = \frac{8\pi}{3}n,
\end{align}
in this limit. Meanwhile, if we assume $|\Delta_{\bf k}| \propto k$ near the Fermi surface, we have
\begin{align}
\int k^3 dk \frac{d_\parallel}{d^3}\frac{\partial d_z}{\partial k}\frac{\partial d_\parallel}{\partial k} = \int k dk \frac{d_\parallel^2}{d^3}2(d_z + \mu)= k_f^2 \int_{-\mu}^\infty d\xi \frac{\Delta^2}{(\xi^2 + \Delta^2)^{3/2}}= 2k_f^2 = 8\pi n.
\end{align}

At the quantum critical point, the ratio of the phonon Hall viscosity to the gravitational Hall viscosity becomes different from what we have for the weak-pairing limit. We can see this from
\begin{align}
\int k dk (1-\hat{d}_z) =& \int k dk \frac{d_\parallel^2}{d(d+d_z)}= \frac{1}{2}\int dk^2 \frac{\hat{\Delta}^2 k^2}{d(d+d_z)}= \frac{1}{4}\left(\frac{2m^*}{\hbar^2}\right)^2 \int_0^\infty d\xi \frac{\hat{\Delta}^2}{\xi + m^*\hat{\Delta}^2/\hbar^2}=4\pi n,\nonumber\\
\int k^2 dk \frac{d_z d_\parallel^2}{d^5}\left(d_z\frac{\partial d_z}{\partial k}+d_\parallel\frac{\partial d_\parallel}{\partial k}\right) =& \int k dk \frac{d_z d_\parallel^2}{d^5}(2d_z^2 + d_\parallel^2)= \left(\frac{2m^*}{\hbar^2}\right)^2 \int_0^\infty d\xi \frac{\hat{\Delta}^2}{\xi + \frac{2m^*\hat{\Delta}^2}{\hbar^2}} \frac{\sqrt{\xi}(\xi + \frac{m^*\hat{\Delta}^2}{\hbar^2})}{(\xi + \frac{2m^*\hat{\Delta}^2}{\hbar^2})^{5/2}}>\frac{8\pi}{3}n,\nonumber\\
\int k^3 dk \frac{d_\parallel}{d^3}\frac{\partial d_z}{\partial k}\frac{\partial d_\parallel}{\partial k} =& \int k dk \frac{d_\parallel^2}{d^3}2d_z=  \left(\frac{2m^*}{\hbar^2}\right)^2 \int_0^\infty d\xi \frac{\hat{\Delta}^2}{\xi + \frac{2m^*\hat{\Delta}^2}{\hbar^2}} \sqrt{\frac{\xi}{\xi + \frac{2m^*\hat{\Delta}^2}{\hbar^2}}}>8\pi n,
\end{align}
where $d_\parallel = \hat{\Delta}k$. We also have
\begin{equation}
4\pi n < \int k dk (1-\hat{d}_z^3) = \int k dk (1-\hat{d}_z)(1+\hat{d}_z+\hat{d}_z^2) < 12\pi n.
\end{equation}
In the strong-pairing limit, we do not find asymptotic limit to the ratio of integrals to $n$ that is independent of cutoff.
\end{widetext}

%\bibliography{hallViscBib,publst,wencross,all,biblio,TI}

%merlin.mbs 2010-03-15 4.21a (PWD, AO, DPC)
%Control: key (0)
%Control: author (8) initials jnrlst
%Control: editor formatted (1) identically to author
%Control: production of article title (-1) disabled
%Control: page (0) single
%Control: year (1) truncated
%Control: production of eprint (0) enabled
%

\end{document}